\documentclass[preprint,11pt]{elsarticle}
\usepackage[margin=2.5cm]{geometry}

\usepackage{graphicx}

\usepackage{mathtools}
\usepackage{booktabs,multirow,array}
\usepackage{makecell}

\usepackage{subfig}
\usepackage{rotating}
\usepackage{url}
\usepackage{lineno}
\usepackage[dvipsnames]{xcolor}

\journal{ISPRS Journal of Photogrammetry and Remote Sensing}

\begin{document}

\begin{frontmatter}



\title{A Framework for SAR-Optical Stereogrammetry over Urban Areas}


\author[1]{Hossein Bagheri}
\author[1]{Michael Schmitt}
\author[2]{Pablo d'Angelo}
\author[1,2]{Xiao Xiang Zhu}

\address[1]{Signal Processing in Earth Observation, Technical University of Munich, Munich, Germany}
\address[2]{Remote Sensing Technology Institute, German Aerospace Center, Oberpfaffenhofen, Wessling, Germany}

\begin{abstract}
\textcolor{Red}{This is the pre-acceptance version, to read the final version, please go to ISPRS Journal of Photogrammetry and Remote Sensing on ScienceDirect}. Currently, numerous remote sensing satellites provide a huge volume of diverse earth observation data. As these data show different features regarding resolution, accuracy, coverage, and spectral imaging ability, fusion techniques are required to integrate the different properties of each sensor and produce useful information. For example, synthetic aperture radar (SAR) data can be fused with optical imagery to produce 3D information using stereogrammetric methods. The main focus of this study is to investigate the possibility of applying a stereogrammetry pipeline to very-high-resolution (VHR) SAR-optical image pairs. For this purpose, the applicability of semi-global matching is investigated in this unconventional multi-sensor setting. To support the image matching by reducing the search space and accelerating the identification of correct, reliable matches, the possibility of establishing an epipolarity constraint for VHR SAR-optical image pairs is investigated as well. In addition, it is shown that the absolute geolocation accuracy of VHR optical imagery with respect to VHR SAR imagery such as provided by TerraSAR-X can be improved by a multi-sensor block adjustment formulation based on rational polynomial coefficients. Finally, the feasiblity of generating point clouds with a median accuracy of about 2m is demonstrated and confirms the potential of 3D reconstruction from SAR-optical image pairs over urban areas.
\end{abstract}

\begin{keyword}
Epipolarity constraint \sep Multi-sensor block adjustment \sep Dense image matching \sep 3D reconstruction \sep SAR-optical stereogrammetry


\end{keyword}

\end{frontmatter}


\section{Introduction}

Three-dimensional reconstruction from remote sensing data has a range of applications across different fields, such as urban 3D modeling and management, environmental studies, and geographic information systems. Manifold high-resolution sensors in space provide the possibility of reconstructing natural and man-made landscapes over large-scale areas. Conventionally, 3D reconstruction in remote sensing is either based on exploiting phase information provided by interferometric SAR, or on space intersection in the frame of photogrammetry with optical images or radargrammetry with SAR image pairs. In all these stereogrammetric approaches, at least two overlapping images are required to extract 3D spatial information. Both photogrammetry and radargrammetry, however, suffer from several drawbacks. Photogrammetry using high-resolution optical imagery is limited by relatively poor absolute localization accuracy and cloud effects, whereas radargrammetry suffers from the difficulty of image matching for severely different oblique viewing angles. 

On the other hand, the huge archives of high-resolution SAR images provided by satellites such as TerraSAR-X and the regular availability of new data alongside archives of high-resolution optical imagery provided by sensors such as WorldView provide a great opportunity to investigate data fusion pipelines for producing 3D spatial information \cite{Schmitt2016a}. 
As relatively few studies have dealt with 3D reconstruction from SAR-optical image pairs \cite{Bloom1988,raggam1994,6690226}, there has been no investigation into the feasibility of a dense multi-sensor stereo pipeline as known from photogrammetric computer vision yet. This paper investigates the possibility of implementing such a pipeline, and describes all processing steps required for 3D reconstruction from very-high-resolution (VHR) SAR-optical image pairs.  

In detail, this paper discusses both an epipolarity constraint and a bundle adjustment formulation for SAR-optical multi-sensor stereogrammetry first. Regarding the complicated radiometric relationship between SAR and optical imagery, the epipolarity constraint accelerates the matching process and helps to identify reliable and correct conjugate points \cite{Morgan2004EpipolarRO,988771}. For this objective, we first demonstrate the existence of an epipolarity constraint for SAR-optical imagery by reconstructing the rigorous geometry models of SAR and optical sensors using both collinearity and range-Doppler relationships. We prove that a SAR-optical epipolarity constraint can be rigorously modeled using the sensor geometries. Subsequently, rational polynomial coefficients (RPCs) are fitted to the SAR sensor geometry to ease further processing steps. Consequently, epipolar curves can be established using projection and back-projection from SAR imagery to terrain and then from terrain to optical imagery using RPCs. In addition, the RPCs ease the formulation of multi-sensor block adjustment for SAR-optical imagery.

The block adjustment is used to align the optical imagery with respect to the SAR data. Generally, the absolute geolocalization accuracy of optical satellite imagery is lower than that of modern SAR sensors. Evaluations show that the absolute accuracy of geopositioning using TerraSAR-X imagery is within a single resolution cell in both the azimuth and range directions, and can even go down to the cm-level \cite{5570983}. In contrast, the absolute accuracy of geolocalization using basic WorldView-2 products is generally no better than 3m \cite{digitalglobe}. Consequently, the block adjustment propagates the high geometrical accuracy of SAR data into the final 3D product, thus avoiding the need for external control points. 

The main stage of SAR-optical stereogrammetry, however, is a dense matching algorithm for 3D reconstruction \cite{Bagheri2018}. In this study, we use the semi-global matching (SGM) method, which incorporates both mutual information and census, as well as their weighted sum as cost functions, in its core.
 
The remainder of this paper is organized as follows. First, the modeling of SAR sensor geometries with RPCs is explained in Section \ref{RPC}. After briefly introducing the epipolarity constraint and its benefits, a mathematical proof of this constraint for SAR-optical image pairs is presented in Section \ref{Epi}. In Section \ref{blockAdj}, the application of multi-sensor block adjustment using RPCs for SAR-optical image pairs is introduced.
The principle of the SGM algorithm is recapitulated in Section \ref{denseMatch}. Section \ref{result} summarizes experiments and results of our implementation of the SAR-optical stereogrammetry workflow for TerraSAR-X/WorldView-2 image pairs over two urban study areas. Based on these results, the feasibility of stereogrammetric 3D reconstruction from SAR-optical image pairs over urban areas, as well as its advantages and limitations, are discussed in Section \ref{sec.dis}. Finally, Section \ref{sec.conclusion} presents the conclusions to this study.

\section{SAR-Optical Stereogrammetry}\label{framework} 

\begin{figure*}[ht!]
	\begin{center}
		\includegraphics[width=1\textwidth]{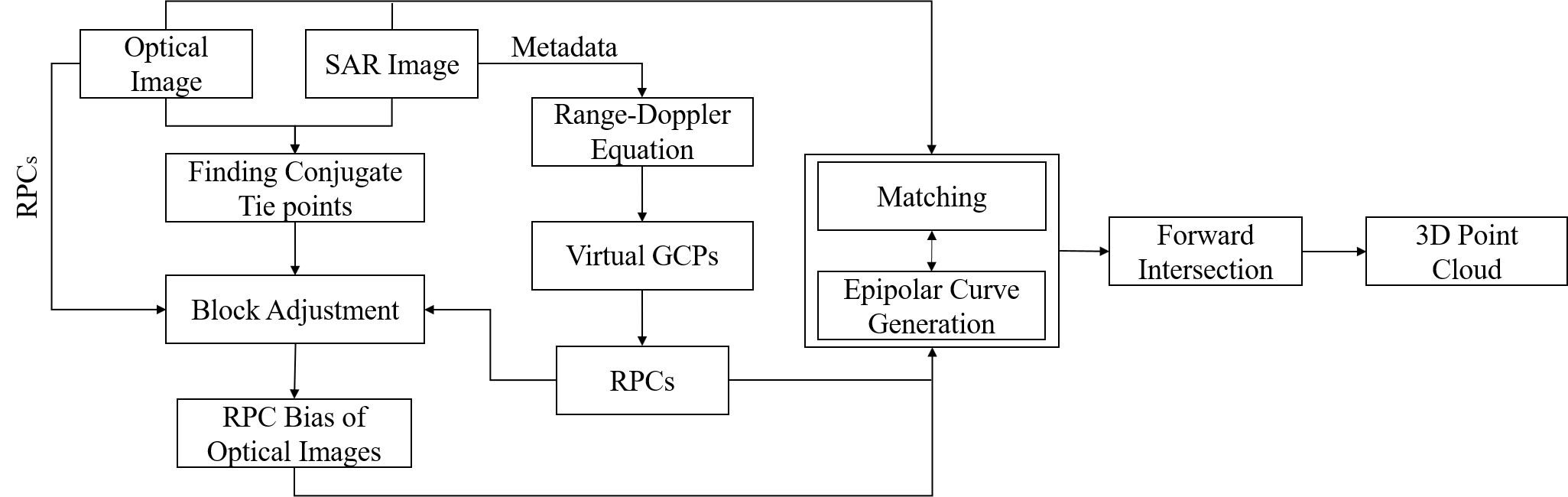}
		\caption{Framework for stereogrammetric 3D reconstruction from SAR-optical image pairs}
		\label{frame}
	\end{center}
\end{figure*}

Figure \ref{frame} shows the general framework of SAR-optical stereogrammetric 3D reconstruction. Similar to optical stereogrammetry, one grayscale optical image and one amplitude SAR image form a stereo image pair that can be processed by suitable matching methods to find all possible conjugate pixels. However, some important pre-processing steps are required before the matching and 3D reconstruction. Currently, most VHR optical images are delivered using RPCs. Thus, the primary step in the SAR-optical stereogrammetry framework is to estimate the RPCs for SAR imagery as well. This process homogenizes the geometry models of both sensors and simplifies the subsequent processes of SAR-optical block adjustment and establishing an epipolarity constraint. The next phase is to carry out multi-sensor block adjustment to align the optical image to the SAR image. This rectifies the RPCs of the optical imagery with respect to the SAR imagery, thus improving the absolute geolocalization of the optical imagery and correcting the positions of the epipolar curves on the optical imagery. A disparity map is then produced in the frame of the reference image via a dense image matching algorithm such as SGM. From this map, the 3D positions of the points can be determined by reconstructing the geometry of the SAR and optical imagery for a particular exposure. However, the success of the aforementioned framework relies on the possibility of establishing an epipolarity constraint for SAR-optical image pairs. Thus, the existence of the epipolarity constraint for SAR-optical image pairs must be investigated. In the following, the details of each step of the SAR-optical stereogrammetry framework are explained and the potential of using an epipolarity constraint for SAR-optical image pairs is investigated.          

\subsection{Preparation: RPCs for SAR Imagery}\label{RPC}
RPCs are a well-established substitute for the rigorously derived optical imaging model. They are widely used for different purposes such as epipolar curve reconstruction \cite{oh2010piecewise}, block adjustment \cite{grodecki2003block}, space resection-intersection and 3D reconstruction \cite{FRASER2006182,li2007integration,tao20023d,doi:10.1080/01431160310001618392,toutin2006comparison} or image rectification \cite{doi:10.1080/07038992.2001.10854900}. The relation between the image space and the geographic reference system is created by the rational functions \cite{Grodecki2004}

\begin{equation}
c= \dfrac{P_{1}(\lambda,\phi, h)}{P_{2}(\lambda,\phi, h)}=f(\lambda,\phi, h) \label{eq.RPCa} 
\end{equation}
and 	
\begin{equation}
r= \dfrac{P_{3}(\lambda,\phi, h)}{P_{4}(\lambda,\phi, h)}=g(\lambda,\phi, h), \label{eq.RPCb}
\end{equation} 
where $r$, $c$ are normalized image coordinates, i.e. normalized rows and columns of points in the scene and $\phi$, $\lambda$, and $h$ denote the normalized latitude, longitude, and height of the respective ground point. The relationship between normalized and un-normalized coordinates is given by \cite{tao2001comprehensive}

\begin{equation}
X=\dfrac{X_{u}-X_{o}}{S_{x}}, \label{eq.normal} 
\end{equation}
where $ X $ is the normalized coordinate, $ X_{u} $ is the un-normalized value of the coordinate, and $ X_{o} $, $ S_{x} $ are the offset and scale factors, respectively.   

In equations (\ref{eq.RPCa}) and (\ref{eq.RPCb}), $P_{i}$ $(i=1,...,4)$ are $n$-order polynomial functions that are used to model the relationship between the image space and the reference system. They can be written as

\begin{equation} 
\begin{split}
P_i= a_{i,0} +a_{i,1}h +a_{i,2}\phi +a_{i,3}\lambda \\
+a_{i,4}h\phi +a_{i,5}h\lambda +a_{i,6}\phi\lambda +a_{i,7}h^2 +a_{i,8}\phi^2 +a_{i,9}\lambda^2 \\
+a_{i,10}h\phi\lambda +a_{i,11}h^{2}\phi +a_{i,12}h^{2}\lambda +a_{i,13}\phi^{2}h +a_{i,14}\phi^{^2}\lambda \\
+a_{i,15}h\lambda^{^2} +a_{i,16}\phi\lambda^{^2} +a_{i,17}h^{3} +a_{i,18}\phi^{^3} +a_{i,19}\lambda^{^3},
\end{split}
\end{equation}
where $ a_{i,n} \:  (n=0,1,...,19)$ are the polynomial coefficients.

For projection from the image space to terrain, the inverse form of the rational function models is used:
\begin{equation}
\lambda= \dfrac{P_{5}(c,r, h)}{P_{6}(c,r, h)}=f^{\prime}(c,r, h) \label{eq.invRPCa} 
\end{equation}
and 	
\begin{equation}
\phi= \dfrac{P_{7}(c,r, h)}{P_{8}(c,r, h)}=g^{\prime}(c,r, h) \label{eq.invRPCb}
\end{equation} 
For this task, another set of RPCs for inverse projection as well as the terrain height $h$ is needed.

The main reason for using RPCs is to facilitate the computational process of the subsequent processing tasks. Instead of describing the stereogrammetric intersection with a combination of the range-Doppler model for the SAR image and a push-broom model for the optical image, from a mathematical point of view the RPC formulation homogenizes everything to a comparably simple joint model. However, fitting RPCs to a sensor model is challenging in its own way and demands sufficient, well-distributed control points. RPCs are usually calculated with either a terrain-independent or terrain-dependent approach \cite{tao2001comprehensive}. In the terrain-dependent approach, accurate Ground Control Points (GCPs) are used to estimate the RPCs. Thus, the final accuracy of the RPCs depends on the number, accuracy, and distribution of GCPs. 

\begin{figure}[ht!]
	\begin{center}
		\includegraphics[width=0.5\linewidth]{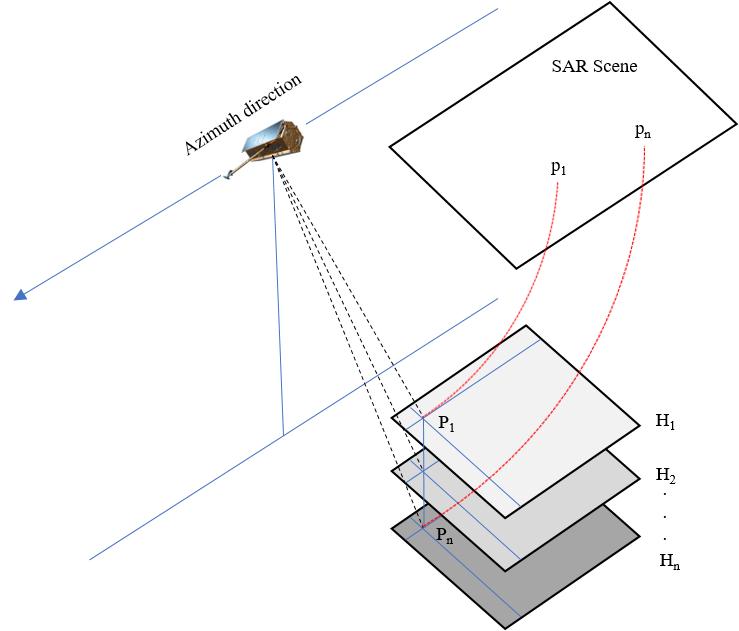}
		\caption{Procedure of estimating RPCs by terrain-independent approach}
		\label{fig.RPC}
	\end{center}
\end{figure}

While the terrain-dependent approach is an expensive way of estimating RPCs (and GCPs may not be available for every study area), the terrain-independent method allows RPCs to be estimated without any GCPs  \cite{tao2001comprehensive}. Instead, a set of virtual GCPs (VGCPs), which are related to the image through the rigorous imaging model of the respective sensor, are used to approximate the RPCs. The VGCPs are arranged in a grid-shape format on planes located at different heights over the study area, such as depicted in Fig \ref{fig.RPC}.	The resulting cube of points is then projected to the image space, and their corresponding image coordinates are determined by reconstructing the rigorous model. The RPCs can subsequently be estimated using a least-squares calculation. Note that, when using higher-order RPCs, the least-squares estimation suffers from an ill-posed configuration that causes the results to deviate from their optimal values. In these circumstances, a regularization approach based on Tikhonov’s method can be employed to obtain acceptable solutions \cite{tikhonov1977methods}. 

The RPCs for optical sensors are usually delivered by vendors alongside the image files. For SAR sensors -- with the exception of the Chinese satellite GaoFen-3 -- however usually only ephemerids and orbital parameters are attached to the data. Therefore, Zhang et al. investigated the generation of RPCs for various SAR sensors based on the terrain-independent approach  \cite{ZHANG2011133}. Their results show that RPCs can be used as substitutes for the range-Doppler equations with acceptable accuracy. In this study, we use this terrain-independent approach for SAR RPC generation.    

\subsection{Epipolarity Constraint for SAR-Optical Image Matching}\label{Epi}
\begin{figure}[ht!]
	\begin{center}
		\includegraphics[width=0.6\linewidth]{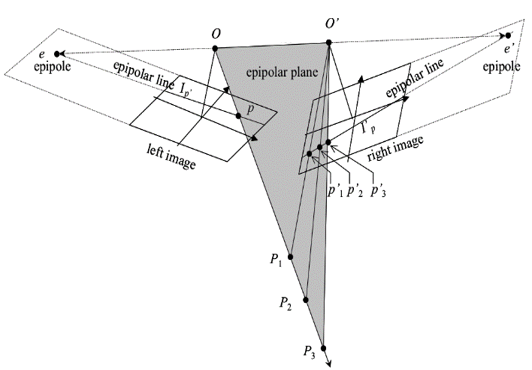}
		\caption{Epipolarity constraint for frame-type camera \cite{Morgan2004EpipolarRO}}
		\label{epi-frame}
	\end{center}
\end{figure}

In most stereogrammetric 3D reconstruction scenarios, the epipolarity constraint facilitates the procedure of image matching by reducing the search space from 2D to 1D \cite{Morgan2004EpipolarRO}. The epipolarity constraint always exists for optical stereo images captured by frame-type cameras that follow a perspective projection \cite{cho1993resampling}. This phenomenon is illustrated in Fig. \ref{epi-frame}.  

For a point $ p $ in the \emph{left-hand} image, the conjugate point in the corresponding \emph{right-hand} image is located on the so-called epipolar line. This epipolar line lies on the plane passing through both image projection centers  ($O,O'$) and the image point $p$. It can also be obtained by changing the depth or height of p in the reference coordinate system. While it is known that epipolar lines exist for images captured from frame-type cameras, straightness cannot be ensured for other sensor types \cite{cho1993resampling}. We thus refer to epipolar curves instead of epipolar lines to express generality in the remainder of this paper.


With respect to remote sensing, several studies have demonstrated that the epipolar curves for scenes acquired by linear array push-broom sensors are not straight \cite{615446,kim2000study}. For example, Kim \cite{kim2000study} used the model developed by Orun and Natarajan \cite{orun1994modified} to prove that the epipolar curves in SPOT scenes looks like hyperbolas. Orun and Natarajan's model assumes that the rotational roll and pitch parameters are constant during the flight, while the yaw can be modeled by quadratic time-dependent polynomials.
Morgan et al. \cite{morgan2004epipolar} demonstrated that the epipolar curves would not be straight even with uniform motion.

For a SAR sensor, the imaging geometry is completely different from that of optical sensors, as data are collected in a side-looking manner based on the range-Doppler geometry \cite{4157311}. However, the possibility of establishing the epipolarity constraint in stereo SAR image pairs has been investigated by Gutjahr et al. \cite{Gutjahr2014TheEC} and Li and Zhang \cite{6509489} for radargrammetric 3D reconstruction. Gutjahr et al. experimentally showed that epipolar curves in SAR image pairs are also not perfectly straight, but can be approximately assumed to be straight for radargrammetric 3D reconstruction tasks through dense matching \cite{Gutjahr2014TheEC}. 

\begin{figure}[ht!]
	\begin{center}
		\includegraphics[width=0.6\linewidth]{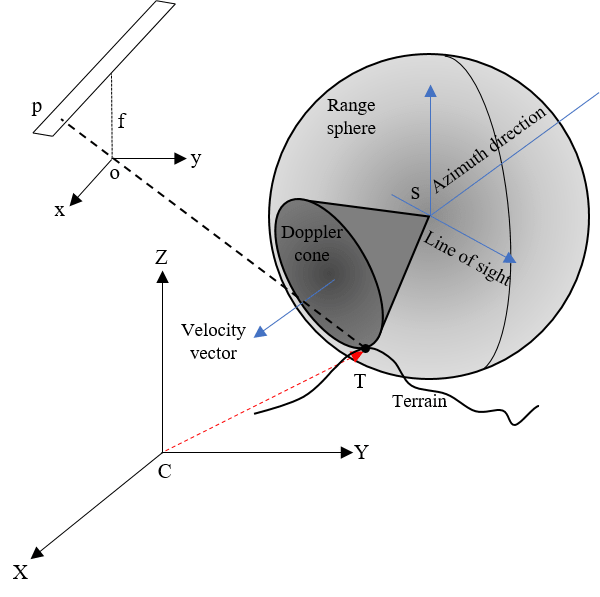}
		\caption{Imaging geometry for configuration of SAR-optical imagery}
		\label{doppler}
	\end{center}
\end{figure}

In this research, we investigate the epipolarity constraint mathematically and experimentally for the unconventional multi-sensor situation of SAR-optical image pairs. In general, epipolar curves in image pairs captured by frame-type cameras (as shown in Fig. \ref{epi-frame})) can be described as \cite{Hartley2004} 

\begin{equation}\label{eq.epi-frame}	   
l_{r}=\mathbf{F}^{T}p^{\prime}
\end{equation}
where $ l_{r} $ refers to the epipolar curve in the right-hand image associated with the image point $ p^{\prime} $ on the left-hand image. $ \mathbf{F} $ is the fundamental matrix, which includes interior and exterior orientation parameters for projecting coordinates between the two images. Similarly, an epipolar curve in the left-hand image can be written as  $ l_{l}=\mathbf{F}p^{\prime \prime} $. For push-broom satellite image pairs, the epipolarity constraint can be verified in a similar way, but linear arrays are substituted for a frame image. Furthermore, the fundamental matrix for push-broom sensors is more complex than that for frame-type sensors. In the following, inspired by the mathematical proof of the epipolarity constraint for stereo optical imagery given in \cite{Morgan2004EpipolarRO} and using the epipolar curve equation presented in (\ref{eq.epi-frame}), a rigorous epipolar model for SAR-optical image pairs acquired by space-borne platforms will be constructed. For this task, the optical image is considered as the left-hand image and the SAR image is the right-hand image. Figure \ref{doppler} shows the configuration of the SAR-optical stereo case. The points $o$ and $s$ mark the positions of the optical linear array push-broom sensor and the SAR sensor, respectively. Using a collinearity condition, a rigorous model for reconstructing the imaging geometry of linear array push-broom sensors can be expressed as  \cite{kratky1988rigorous}

\begin{equation}\label{eq.colli1}	   
\Bigg( 
\begin{tabular}{c}
$x_{l}=0$ \\
$y_{l}$  \\
$f$
\end{tabular}  
\Bigg) = \lambda R_{\omega(t)\phi(t)\kappa(t)} \Bigg( 
\begin{tabular}{c}
$X-X^{o}(t)$ \\
$Y-Y^{o}(t)$ \\
$Z-Z^{o}(t)$
\end{tabular}  
\Bigg), 	        
\end{equation}    

where $(x_{l},y_{l})$ are the coordinates of point $p$ in the linear array coordinate system,  $f$ is the focal length, $(X^{o}(t),Y^{o}(t),Z^{o}(t))$ represents the satellite position at time $t$ in the reference coordinate system, $(X,Y,Z)$ are the ground coordinates of the target point $T$, $\lambda$ is the scale factor, and $ \mathbf{R}_{\omega(t)\phi(t)\kappa(t)}$ is the 3D rotational matrix computed from rotations $\omega(t),\phi(t),\kappa(t)$ along the three dimensions at time $t$. 
Note that the aforementioned rotational and translational components are estimated by time-dependent polynomials.

A rigorous model based on the range-Doppler geometry \cite{curlander1991synthetic} (displayed in Fig. \ref{doppler} as well) can also be applied to the SAR imagery. In this model, the slant-range equation is first used to describe the range sphere as \cite{4157311}:
\begin{equation}\label{eq.range1}	   
R=\|\mathbf{R}_{CT}-\mathbf{R}_{CS}\|
\end{equation}
where $R$ is the slant-range and $\mathbf{R}_{CT}$, $\mathbf{R}_{CS}$ are the target point and SAR sensor position vectors in the reference coordinate system. C refers to the center of the reference coordinate system.

For a given pixel $y_{r}$ in the slant-range SAR scene, equation (\ref{eq.range1}) can be reformulated as
\begin{equation}\label{eq.range2}
\begin{split}	   
R=c \: t=c \: (t_{0}+\dfrac{y_{r}}{2f_{r}}) 
= c \: t_{0} + c \: \dfrac{y_{r}}{2f_{r}}=R_{0}+\Gamma \: y_{r},
\end{split}
\end{equation}
where $R$ is the slant-range of the target point, $c$ is the velocity of light, $t_{0}$, $t$ are the one-way signal transmission times for the first range pixel and range pixel coordinate $y_{r}$, respectively, and $f_{r}$ is the range sampling rate. $R_{0}$ gives the slant-range for the first range pixel and  $\Gamma = \dfrac{c}{2f_{r}}$.

The second equation describes the geometry of the Doppler cone:

\begin{equation}\label{eq.dop}	   
f_{D}=\dfrac{2}{\lambda_{r} R} \mathbf V \cdot  (\mathbf{R}_{CT}-\mathbf{R}_{CS}),
\end{equation}
where $f_{D}$ is the Doppler frequency, $\lambda_{r}$ is the SAR signal wavelength, $\mathbf V$ is the velocity vector and $\cdot$ denotes the inner product operator.

For the epipolarity constraint, we assume that the target point  $ T $ is imaged at time $\tau$ by the push-broom sensor. If we fix the time variable $t$ with $\tau$ and consider the corresponding image coordinate in the linear array coordinate system as $(0,y_{l}, f)$, we can use equation (\ref{eq.colli1}) to back-project from the linear array coordinate system to the terrestrial reference system  as follows:

\begin{equation}\label{eq.colli2}	   
\Bigg( 
\begin{tabular}{c}
$X-X^{o}$ \\
$Y-Y^{o}$ \\
$Z-Z^{o}$
\end{tabular}  
\Bigg) = \lambda^{-1} M_{\omega\phi\kappa} \Bigg( 
\begin{tabular}{c}
$0$ \\
$y_{l}$  \\
$f$
\end{tabular}  
\Bigg) 	        
\end{equation} 
where $\mathbf{M}_{\omega\phi\kappa}=\mathbf{R}_{\omega\phi\kappa}^{T}$.
For imaging time $t$,  all time-dependent parameters are estimated under the constraint $t=\tau$. Thus, for this specific instance, the variable index t is eliminated from equation (\ref{eq.colli1}). By expanding equation (\ref{eq.colli2}) and removing the scale factor effect, we have:
\begin{equation}\label{eq.colli3a}	   
\begin{split}
X-X^{o} =(Z-Z^{o}) \: \dfrac{m_{11} \: 0+m_{12} \: y_{l}+m_{13} \: f}{m_{31} \: 0+m_{32}  \: y_{l}+m_{33} \: f}
=(Z-Z^{o}) \: \dfrac{m_{12} \: y_{l}+m_{13} \: f}{m_{32} \: y_{l}+m_{33} \: f}
\end{split}
\end{equation} 	
\begin{equation}\label{eq.colli3b}	   
\begin{split} 
Y-Y^{o} = (Z-Z^{o}) \: \dfrac{m_{21} \: 0+m_{22} \: y_{l}+m_{23}  \: f}{m_{31} \: 0+m_{32} \: y_{l}+m_{33} \: f}= (Z-Z^{o})\dfrac{m_{22} \: y_{l}+m_{23} \: f}{m_{32} \: y_{l}+m_{33} \: f}
\end{split}          
\end{equation} 
where $m_{ij}$ are elements of matrix $\mathbf{M}_{\omega\phi\kappa}$.

If the velocity vector of the SAR sensor is computed in the zero-Doppler frequency transition, we can reformulate equation (\ref{eq.dop}) as:

\begin{equation}
\begin{split}
\label{eq.dopEx1}
V_{x}(t)(X-X^{s}(t))+V_{y}(t)(Y-Y^{s}(t))+V_{z}(t)(Z-Z^{s}(t))=0
\end{split}
\end{equation}
where $(V_{x}(t),V_{y}(t),V_{z}(t))$ are the components of velocity vector $\mathbf V$. Hence: 

\begin{equation}
\begin{split}
\label{eq.dopEx2}	   
V_{x}(t)X-V_{x}(t)X^{s}(t)+V_{y}(t)Y-V_{y}(t)Y^{s}(t)+V_{z}(t)Z-V_{z}(t)Z^{s}(t)=0
\end{split}
\end{equation}
From equations (\ref{eq.colli3a}) and (\ref{eq.colli3b}),  we can derive. 

\begin{equation}
\label{eq.colli4a}
X- \big (\frac{m_{12} \: y_{l}+m_{13} \: f}{m_{32} \: y_{l}+m_{33} \: f} \big ) \: Z=X^{o}-\big (\frac{m_{12} \: y_{l}+m_{13} \: f}{m_{32} \: y_{l}+m_{33} \: f} \big ) \: Z^{o} 
\end{equation}
\begin{equation}
\label{eq.colli4b}
Y- \big ( \frac{m_{22}  \: y_{l}+m_{23} \: f}{m_{32} \: y_{l}+m_{33} \: f} \big ) \: Z=Y^{o}-  \big ( \frac{m_{22} \: y_{l}+m_{23} \: f}{m_{32} \: y_{l}+m_{33} \: f} \big ) \: Z^{o} 
\end{equation}
Multiplying both sides of the equations (\ref{eq.colli4a}) and (\ref{eq.colli4b}) by $-V_{x}(t)$ and $-V_{y}(t)$, respectively, and then combining them with equation (\ref{eq.dopEx2}), Z can be calculated as follows:
\begin{equation}
\begin{split}
\label{eq.Z1}
Z= \dfrac{(m_{32} \:  y_{l}+m_{33} \: f) \:  \big [V_{x}(t)(X^{s}(t)-X^{o})+V_{y}(t)(Y^{s}(t)-Y^{o})+V_{z}(t)Z^{s}(t) \big]}{(m_{12} \: y_{l}+m_{13} \: f) \:  V_{x}(t)+(m_{22} \: y_{l}+m_{23} \: f) \:  V_{y}(t)+(m_{32} \: y_{l}+m_{33} \: f) \: V_{z}(t)}\\+ \dfrac{ \big [(m_{12}  \: y_{l}+m_{13} \: f) \:  V_{x}(t)+(m_{22} \: y_{l}+m_{23} \: f) \:  V_{y}(t) \big] Z^{o}}{(m_{12} \: y_{l}+m_{13} \: f) \:  V_{x}(t)+(m_{22} \: y_{l}+m_{23} \: f) \:  V_{y}(t)+(m_{32} \: y_{l}+m_{33} \: f) \:  V_{z}(t)} 
\end{split}
\end{equation}

Changing the position of target point $T$ in the $Z$ direction is equivalent to changing the corresponding image coordinates on the epipolar curve. Consequently, the determination of image coordinates in the SAR scene can be realized by tracking the sensor positions in the respective instances.
In fact, in spite of fixing the position of the target point for the optical sensor at time $\tau$, the location components $(X^{s}(t),Y^{s}(t),Z^{s}(t))$ and the velocity components of the SAR sensor $(V_{x}(t),V_{y}(t),V_{z}(t))$ are time-dependent and can be estimated for each instant using time-dependent polynomials. For the sake of simplicity, the SAR sensor trajectory can be approximated by a linear motion model. Thus, the velocity components $(V_{x}(t),V_{y}(t),V_{z}(t))$ will remain constant with time at $(X^{s}(t),Y^{s}(t),Z^{s}(t))$ — i.e., the acceleration is 0 —  and the location components $(X^{s}(t),Y^{s}(t),Z^{s}(t))$ can be calculated using linear time-dependent functions:

\begin{equation}\label{eq.position}	   
\begin{tabular}{c}
$X^{s}(t_{a}) =X^{s}_{0}+V_{x} \: t_{a}$ \\
$Y^{s}(t_{a}) =Y^{s}_{0}+V_{y} \: t_{a}$ \\
$Z^{s}(t_{a}) =Z^{s}_{0}+V_{z} \: t_{a}$ 
\end{tabular}           
\end{equation}    	
where $(X^{s}_{0}, Y^{s}_{0},Z^{s}_{0})$ is the position of the SAR sensor at initial time $ t_{0} $. The time $t_{a}$ is the azimuthal time, which can be expressed according to the line coordinate $x_{r}$ in the SAR scene as
\begin{equation}\label{eq.time}	   
t_{a}=\dfrac{x_{r}}{{PRF}}=k \: x_{r} \:  ,
\end{equation}
where ${PRF}$ is the pulse repetition frequency in Hz.

Substituting the parameters expressed in equations (\ref{eq.position}) and (\ref{eq.time}) into (\ref{eq.Z1}), we obtain:

\begin{equation}\label{eq.Z2}
\begin{split}
Z= \dfrac{(m_{32} \: y_{l}+m_{33} \: f) \: \big [V_{x}(X^{s}_{0}-X^{o})+V_{y}(Y^{s}_{0}-Y^{o})+V_{z}Z^{s}_{0} \big]}{(m_{12} \: y_{l}+m_{13} \: f) \: V_{x}+(m_{22} \: y_{l}+m_{23} \: f) \: V_{y}+(m_{32} \: y_{l}+m_{33} \: f) \: V_{z}} \\ +\dfrac{ \big [ (m_{12} \: y_{l}+m_{13} \: f) \: V_{x}+(m_{22} \: y_{l}+m_{23} \: f) \: V_{y} \big ] Z^{o} } {(m_{12} \: y_{l}+m_{13} \: f) \: V_{x}+(m_{22} \: y_{l}+m_{23} \: f) \: V_{y}+(m_{32} \: y_{l}+m_{33} \: f) \: V_{z}}\\+k \: x_{r} \: \big ( \dfrac{V_{x}^{2}+V_{y}^{2}+V_{z}^{2}}{(m_{12} \: y_{l}+m_{13} \: f) \: V_{x}+(m_{22} \: y_{l}+m_{23} \: f) \: V_{y}+(m_{32} \: y_{l}+m_{33} \: f) \: V_{z}} \big )
\end{split}
\end{equation}

Equation (\ref{eq.Z2}) can be simplified to: 

\begin{equation}\label{eq.Z3}
Z= c_{0}+c_{1} \: x_{r}
\end{equation} 
 and substituting (\ref{eq.Z3}) into (\ref{eq.colli4a}) and (\ref{eq.colli4b}) gives:

\begin{equation}\label{eq.XY1a}
\begin{split}
X=X^{o}- \big ( \frac{m_{12} \: y_{l}+m_{13} \: f}{m_{32} \: y_{l}+m_{33} \: f} \big ) \: Z^{o}+ \big ( \frac{m_{12} \: y_{l}+m_{13} \: f}{m_{32} \: y_{l}+m_{33} \: f} \big ) (c_{0}+c_{1}  \: x_{r})\\=X^{o}-\big ( \frac{m_{12} \: y_{l}+m_{13} \: f}{m_{32} \: y_{l}+m_{33} \: f} \big ) \: Z^{o}+ \big ( \frac{m_{12} \: y_{l}+m_{13} \: f}{m_{32} \: y_{l}+m_{33}  \: f} \big ) \: c_{0}+ \big ( \frac{m_{12} \: y_{l}+m_{13} \: f}{m_{32} \: y_{l}+m_{33} \: f} \big ) \: c_{1} \: x_{r} 
\end{split}
\end{equation}

	\begin{equation}\label{eq.XY1b}
\begin{split}
Y=Y^{o}-\big ( \frac{m_{22} \: y_{l}+m_{23} \: f}{m_{32} \: y_{l}+m_{33} \: f}\big ) \: Z^{o}+\big ( \frac{m_{22} \: y_{l}+m_{23} \: f}{m_{32} \: y_{l}+m_{33} \: f}\big ) \: (c_{0} +c_{1} \: x_{r})=\\Y^{o}-\big ( \frac{m_{22} \: y_{l}+m_{23} \: f}{m_{32} \: y_{l}+m_{33} \: f} \big ) \: Z^{o}+\big ( \frac{m_{22} \: y_{l}+m_{23} \: f}{m_{32} \: y_{l}+m_{33} \: f}\big ) \: c_{0}+\big ( \frac{m_{22} \: y_{l}+m_{23} \: f}{m_{32} \: y_{l}+m_{33} \: f} \big ) \: c_{1} \: x_{r} 
\end{split}
\end{equation} 

Equations (\ref{eq.XY1a}) and (\ref{eq.XY1b}) can be written in the form:
\begin{equation}
X= a_{0}+a_{1} \: x_{r} \label{eq.XY2a} 	
\end{equation} 
\begin{equation}
Y= b_{0}+b_{1} \: x_{r} \label{eq.XY2b}
\end{equation} 

The slant-range represented by equation (\ref{eq.range2}) can be reformulated as:

\begin{equation}\label{eq.range3}
(X-X^{s})^{2}+(Y-Y^{s})^{2}+(Z-Z^{s})^{2}=(R_{0}+\Gamma \: y_{r})^{2}
\end{equation} 

Substituting  (\ref{eq.XY2a}), (\ref{eq.XY2b}), (\ref{eq.Z3}) and (\ref{eq.position}) into (\ref{eq.range3}), we have:

\begin{equation}\label{eq.epi1}
\begin{split}
(a_{0}+a_{1} \: x_{r}-X^{s}_{0}-V_{x} \: k \: x_{r})^{2}+(b_{0}+b_{1} \: x_{r}-Y^{s}_{0}-V_{y} \: k \: x_{r})^{2}\\+(c_{0}+c_{1} \: x_{r}-Z^{s}_{0}-V_{z} \: k \: x_{r})^{2}=(R_{0}+\Gamma \: y_{r})^{2}
\end{split}	
\end{equation}   

For simplicity, setting  $A_{0}=a_{0}-X^{s}_{0}$, $A_{1}=a_{1}-V_{x}\:k$, 
$B_{0}=b_{0}-Y^{s}_{0}$, $B_{1}=b_{1}-V_{y}\:k$, $C_{0}=c_{0}-Z^{s}_{0}$, $C_{1}=c_{1}-V_{z}\:k$
gives:
\begin{equation}\label{eq.epi2}
\begin{split}
(A_{0}+A_{1}\:x_{r})^{2}+(B_{0}+B_{1}\:x_{r})^{2}+(C_{0}+C_{1}\:x_{r})^{2}=(R_{0}+\Gamma \:y_{r})^{2}
\end{split}	
\end{equation}   
which can be expanded to yield:

\begin{equation}\label{eq.epi3}
\begin{split}
(A_{0}^{2}+B_{0}^{2}+C_{0}^{2})+2(A_{0}\:A_{1}+B_{0}\:B_{1}+C_{0}\:C_{1})\:x_{r}\\+(A_{1}^{2}+B_{1}^{2}+C_{1}^{2})\:x_{r}^{2}=(R_{0}+\Gamma \:y_{r})^{2}
\end{split}	
\end{equation} 

If $A_{0}^{2}+B_{0}^{2}+C_{0}^{2}=F_{0}$ , $2(A_{0}A_{1}+B_{0}B_{1}+C_{0}C_{1})=F_{1}$, and $A_{1}^{2}+B_{1}^{2}+C_{1}^{2} =F_{2}$,
this can be rewritten as:

\begin{equation}\label{eq.epif}
\Gamma \: y_{r}=\sqrt {F_{2} \:x_{r}^{2}+F_{1}\:x_{r}+F_{0}}-R_{0}
\end{equation} 

Equation (\ref{eq.epif}) is a general rigorous model representing the epipolarity constraint for SAR-optical image pairs based on their imaging parameters contained in $F_0, F_1, F_2, \Gamma$ and $R_0$. 
This shows that an epipolarity-like constraint can be established for SAR-optical image pairs. However, the non-linear relation between $y_{r}$ and $x_{r}$ in equation (\ref{eq.epif}) shows that SAR-optical epipolar curves are not straight, even under the assumption of linear motion for the SAR system. In Section \ref{subsec.dis-epi}, this epipolarity constraint will be experimentally investigated for an RPC-based imaging model.

\subsection{SAR-Optical Multi-Sensor Block Adjustment}\label{blockAdj}
As illustrated in Fig. \ref{frame}, the main step before implementing dense image matching is to align the optical image to the SAR image. This process is performed using a multi-sensor block adjustment which is based on RPCs instead of rigorous sensor models as proposed in \cite{grodecki2003block}. The block adjustment process improves the relative orientation between both images fixed to the more accurate SAR image orientation parameters. Through the block adjustment, the bias components induced by attitude, ephemeris, and drift errors in the optical image are compensated \cite{dlr78910}.

The main bias compensation for the RPCs of the optical image involves translating the locations of the epipolar curves to accurate positions using the SAR geopositioning accuracy. Generally, designing an appropriate function for modeling the existing bias in the RPCs given by the optical image depends on the sensor properties \cite{TONG2010218}, but for most sensors an affine model can be applied \cite{fraser2005bias}. Even for the current generation of VHR linear push-broom array sensors such as WorldView-2, employing only the shift parameters will be sufficient. The affine model for RPC bias compensation can be formulated as

\begin{equation}\label{eq.BA}
\begin{array}{lcl} 
\Delta x&=&m_{0}+m_{1}\:x_{o}+m_{2}\:y_{o} \\
\Delta y&=&n_{0}+n_{1}\:x_{o}+n_{2}\:y_{o},
\end{array}
\end{equation}   
where $x_{o}$,$y_{o}$ represent column and row of tie points in the optical images and $m_{i}$ and $n_{i}$ $ (i=0,1,2) $ are unknown affine parameters to be estimated through the block adjustment procedure. Note that tie points are the common points between the SAR and optical images, and can be obtained by manual or automatic sparse matching between two images. Since the automation of the tie point generation process is not the focus of this study, we refer the reader to possible solutions described in \cite{5340570, 6049732, Merkle2017}.  

The geographic coordinates of the tie points in the SAR image are calculated by the inverse rational functions computed for the SAR imagery as described in Section~\ref{RPC}:

\begin{equation}
\lambda^{i}=f_{s}^{'}(x_{s}^{i},y_{s}^{i}, H) \label{eq.iRPCa} 
\end{equation}
and 	
\begin{equation}
\phi^{i}= g_{s}^{'}(x_{s}^{i},y_{s}^{i}, H) \label{eq.iRPCb}
\end{equation}
where  $ \lambda^{i} $ and $ \phi^{i} $ are the normalized longitude and latitude of tie point $ i $ with  normalized image coordinates $ x_{s}^{i} $ and $ y_{s}^{i} $ (index $s$ denotes the SAR scene), and here, $H$ is a constant, e.g., the mean height of the study area. $ f_{s}^{'} $ and $ g_{s}^{'} $ are inverse rational functions computed for the SAR sensor to project the tie points from the SAR image to the reference system. The output is a collection of GCPs that can be applied for the RPC rectification of the optical imagery. The resulting GCPs are then projected by the rational function associated with the optical images to give the image coordinates of the GCPs:

\begin{equation}
c_{o}^{i} =f_{o}(\lambda^{i},\phi^{i}, H) \label{eq.RPCa1} 
\end{equation}
and 	
\begin{equation}
r_{o}^{i} =g_{o}(\lambda^{i},\phi^{i}, H) \label{eq.RPCb1}
\end{equation} 
where $ c_{o}^{i} $, $ r_{o}^{i} $ are the normalized image coordinates of tie point $ i $ computed by the forward rational functions of the optical sensor, $ f_{o} $ and $ g_{o} $.

From equations (\ref{eq.BA}), (\ref{eq.RPCa1}), and (\ref{eq.RPCb1}), the primary equations for SAR- optical block adjustment are formed as:

\begin{equation}
x_{o}^{i} = c_{ou}^{i}+\Delta x^{i} +v_{x}^{i} \label{eq.ba.a} 
\end{equation}
and 	
\begin{equation}
y_{o}^{i} =r_{ou}^{i}+\Delta y^{i} +v_{y}^{i} \label{eq.ba.b}
\end{equation} 
where, $ x_{o}^{i} $ and $ y_{o}^{i} $ denote the column and row of tie point $ i $ in the optical scene, and $ c_{ou}^{i} $ and $ r_{ou}^{i} $ are the un-normalized coordinates of the tie point after projection and back-projection using the RPCs. The block adjustment equations can then be written as:

\begin{equation}
F_{x}^{i} = -x_{o}^{i}+c_{ou}^{i}+\Delta x^{i} +v_{x}^{i}=0 \label{eq.ba.a1}, 
\end{equation}
and 	
\begin{equation}
F_{y}^{i} =-y_{o}^{i}+r_{ou}^{i}+\Delta y^{i} +v_{y}^{i}=0 \label{eq.ba.b2}.
\end{equation} 

Finally, through an iterative least-squares adjustment \cite{grodecki2003block}, the unknown parameters $m_{i}$ and $n_{i}$ are estimated and the affine model can be formed. 
This affine model is added to the rational functions of the optical image to improve the geolocation accuracy to that of the SAR image.

\subsection{SGM for Dense Multi-Sensor Image Matching}\label{denseMatch}

The core step in a stereogrammetric 3D reconstruction workflow is the dense image matching algorithm to obtain the disparity map, which can then be transformed into the desired 3D point cloud. Generally, there are two different dense matching rationales that can be used according to whether local or global optimization is more important \cite{Brown2003}. For the case of global optimization, an energy functional consisting of two terms is established to find the optimal disparity map \cite{988771}:

\begin{equation}\label{global}	        
E(d)= E_{data}(d)+\lambda E_{smooth}(d)	        
\end{equation}
where $E_{data}(d)$ is a fidelity term that makes the computed disparity map consistent with the input image pairs, $E_{smooth}(d)$ considers the smoothness condition for the disparity map, and $\lambda$ is a regularization parameter that balances the fidelity and smoothness terms.

For a given image pair, the disparity map is calculated by minimizing the energy functional in (\ref{global}). The main advantage of global dense matching over local matching methods is greater robustness against noise \cite{Brown2003}, although most existing algorithms for global dense image matching have a greater computational cost \cite{Hirschmuller2008}.

For the experiments presented in this paper, we use the well-known SGM method \cite{Hirschmuller2008}, which offers acceptable computational cost and high efficiency, and performs very similarly to global dense image matching.

\subsubsection{Cost Functions used in SGM}\label{subsec.costs}   

In this study, the ability of performing dense image matching for SAR-optical image pairs using SGM is investigated. For this purpose, two different cost functions, namely Mutual Information (MI) and Census, as well as their weighted sum, are examined for the dense matching of high-resolution SAR and optical imagery. Typically, the similarity measures employed in the cost function are either signal-based or feature-based metrics \cite{Hassaballah2016}. Classically, signal-based similarity measures such as Normalized Cross Correlation (NCC) and MI are preferable to feature-based similarity measures when used in dense image matching algorithms because of faster calculation. 

Among the signal-based matching measures, MI was recommended for SGM as it is known to perform well for images with complicated illumination relationship, such as SAR-optical image pairs \cite{5340570}. 

Another similarity measure used in the SGM cost function is Census, which actually acts as a nonparametric transformation. The weighted sum of MI and Census is beneficial for 3D reconstruction in urban areas, especially for reconstructing the footprints of buildings to produce sharper and clearer images  \cite{10.1007/978-3-642-24393-6_14}. The weighted similarity measure can be defined as 
\begin{equation}\label{eq.SM}	        
SM =\alpha \: MI+(1-\alpha) \: \text{Census},
\end{equation} 
where $\alpha$ changes from 0 to 1 to weigh the effect of Census cost in relation to MI.

\subsubsection{SGM Settings for Efficient SAR-Optical Dense Matching}\label{subsec.setting} 
To increase the efficiency of the SGM performance, some important settings for the dense matching of SAR-optical image pairs must be considered. The basic principle of 3D reconstruction by dense matching is to use the epipolarity constraint to limit the search space. Usually, before dense matching, normal images are created by resampling the original images according to epipolar geometry  \cite{Morgan2004EpipolarRO,oh2010piecewise}. In this study, we use the RPC model to realize the SAR-optical epipolar geometry implicitly without the need to generate normal images. This is done by implementing projections and back-projections from the reference image to the ground and back to the corresponding image, respectively, for a specified height range using rational functions. Then, the search for computing disparities can be performed along the thus-created epipolar curves.

In addition, the minimum and maximum disparity values should be selected to restrict the length of the search space along the epipolar curves. In general, there is more flexibility regarding the selection of this disparity interval for optical image pairs than for SAR-optical image pairs, and using unsuitable values will result in more outliers. The minimum and maximum disparity values can be determined using external data such as the SRTM digital elevation model, which is available for most land surfaces around the world \cite{USGS2000}. For sake of exploiting the simplicity offered by comparably flat study scenes, we just add and subtract 20-$ m $ height differences to the mean terrain heights of the study scenes to obtain the disparity thresholds. 

The next setting is to switch off the \textit{minimum region size} option in
the SGM algorithm, which is usually used to decrease the noise level in the stereogrammetric 3D reconstruction of optical image pairs by eliminating isolated patches from the disparity map based on their small size. Experimental results show that, for SAR-optical image pairs, the complex illumination relationship between the images and the different imaging effects (especially for urban areas) make the \textit{minimum region size} criterion useless, as connectivity cannot be ensured in the disparity map.
 
Similar to other dense matching cases, we use the LR (Left-Right) check to investigate binocular half-occlusions \cite{Egnal2002}. This strategy changes the reference images from left to right, and consequently produces two disparity maps that can be checked against each other. To reach sub-pixel accuracy, the disparity in each point is estimated by a quadratic interpolation of neighboring disparities.   

In this study, SGM is implemented at four hierarchy levels and the aggregated cost is calculated along 16 directions around each point.

\section{Experiments and Results}\label{result}

\subsection{Study Areas and Datasets}\label{data}
\begin{figure*} 
	\centering
		\includegraphics[width=1\linewidth]{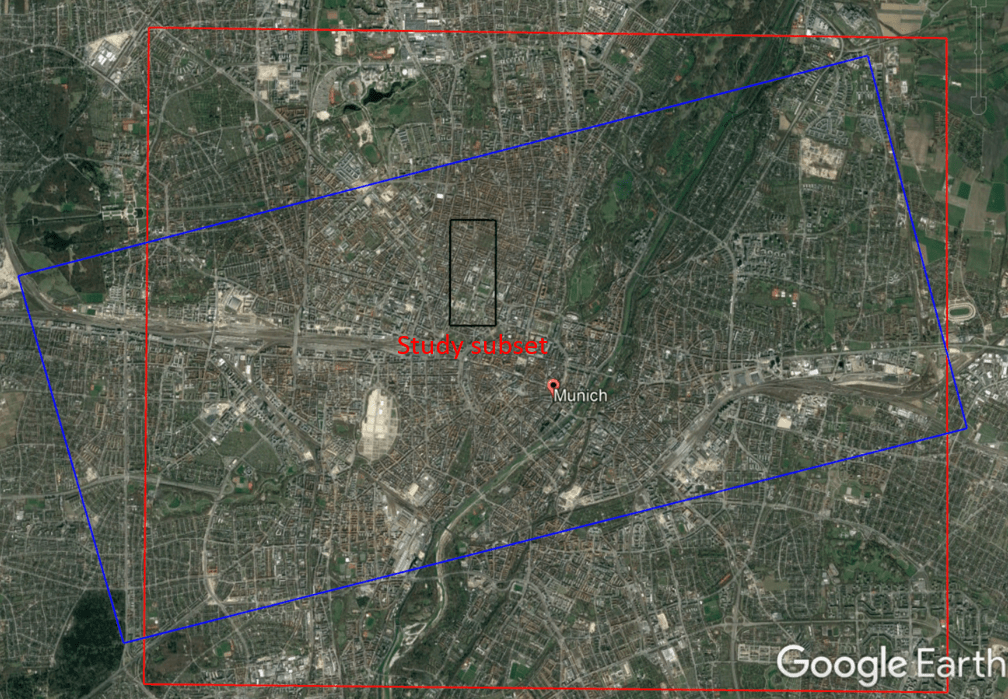}
	\caption{Display the location of SAR-optical image pairs of the Munich study area. The red and blue rectangles identify  areas covered by the WorldView-2 and TerraSAR-X images, respectively,  and the black rectangle displays the study subset selected for stereogrammetrix 3D reconstruction over Munich.}
	\label{fig:studyareaM} 
\end{figure*}

\begin{figure*} 
	\centering
		\includegraphics[width=1\linewidth]{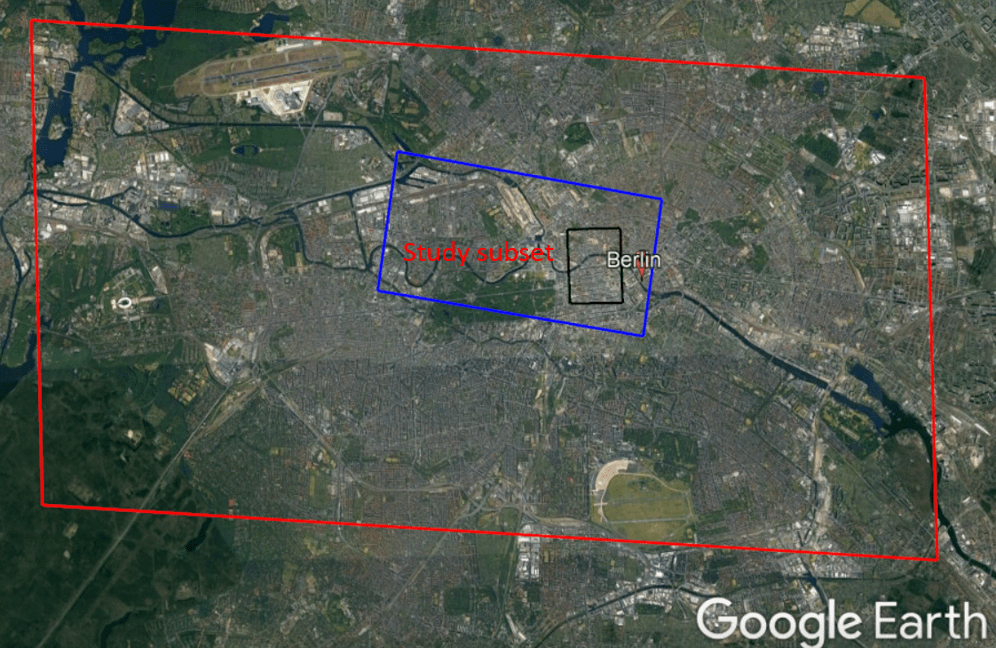}
	\caption{Display the location of SAR-optical image pairs of the Berlin study area. The red and blue rectangles identify  areas covered by the WorldView-2 and TerraSAR-X images, respectively, and the black rectangle displays the study subset selected for stereogrammetrix 3D reconstruction over Berlin.}
	\label{fig:studyareaB} 
\end{figure*}

\begin{figure} 
	\centering
		\includegraphics[width=0.75\linewidth, height=0.6\linewidth]{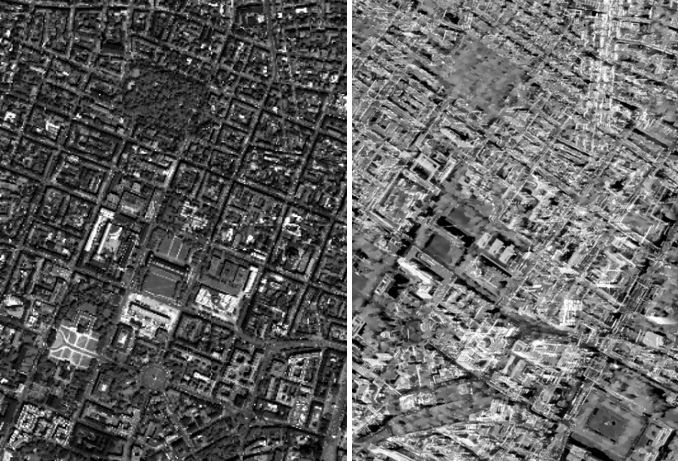}
        \hfill	
	\caption{Display of SAR-optical sub-scenes extracted from Munich study areas (The left-hand image is from WorldView-2, right-hand image is from TerraSAR-X)}
	\label{fig:sub-municharea} 
\end{figure}

\begin{figure} 
	\centering
	\includegraphics[width=0.75\linewidth, height=0.6\linewidth]{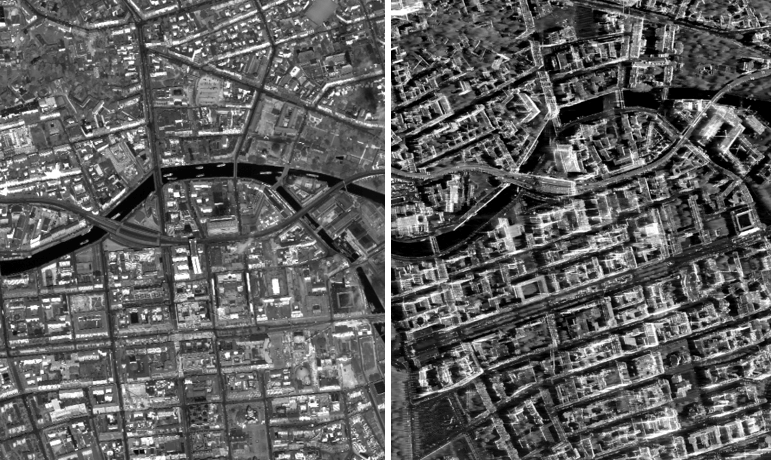}
	\label{fig:sub-Berlin}\\
	\caption{Display of SAR-optical sub-scenes extracted from Berlin study areas (The left-hand image is from WorldView-2, right-hand image is from TerraSAR-X)}
	\label{fig:sub-berlinarea} 
\end{figure}

We selected two study areas, one in Berlin and one in Munich (both located in Germany), to investigate the potential for 3D reconstruction from high-resolution SAR-optical image pairs over urban areas. The locations of TerraSAR-X and WorldView-2 images are displayed in Figs. \ref{fig:studyareaM} and \ref{fig:studyareaB}. The properties of the image pairs for each study area are presented in Table \ref{dataset}. In order to enhance the general image similarity and facilitate the matching process, all images were resampled to 1 $ m $ $\times$ 1 $ m $ pixel spacing and the SAR images were filtered with a non-local speckle filter. After implementing bundle adjustment for both datasets, two sub-scenes (with a size of 1000 $ \times $ 1500 pixels each) from overlapped parts of the study areas were cropped. These sub-scenes are displayed in Figs. \ref{fig:sub-municharea} and \ref{fig:sub-berlinarea}.    

\begin{table*}[h]
	\centering \footnotesize
	\caption{Specifications of the TerraSAR-X and WorldView-2 images used for dense matching}
	\label{dataset}
	\begin{tabular}  {l ccccc}
		 \textbf{Area} &\textbf{Sensor} & \textbf{\thead{Acquisition \\ Mode}} & \textbf{\thead{Off-Nadir \\ Angle ($^\circ$)}} & \textbf{\thead{Ground Pixel \\ Spacing (m)}}&\textbf{ \thead{Acquisition \\ date}}  \\\hline
        \toprule
		\multirow{2}{*}{Munich} 
		&TerraSAR-X	&Spotlight	&22.99& 0.85 $\times$ 0.45& 03.2015 \\ 
		&WorldView-2&Panchromatic	&5.2& 0.5 $\times$ 0.5& 07.2010 \\
        \midrule
		\multirow{2}{*}{Berlin}
		&TerraSAR-X	&Staring Spotlight	&36.11& 0.17 $\times$ 0.45& 04.2016 \\
		&WorldView-2&Panchromatic	&29.1& 0.5 $\times$ 0.5& 05.2013 \\
	\end{tabular}
	
\end{table*}

\subsection{Validation of RPCs to Model SAR Sensor Geometry}\label{subsec.dis-rpc}
As described in Section \ref{RPC}, RPCs can be used as a substitute for the rigorous range-Doppler model, similar to the standard RPCs delivered with optical imagery. This step is performed to simplify the multi-sensor block adjustment and epipolarity constraint construction. The accuracy of the RPCs can be estimated using independent virtual checkpoints that are produced in a similar way to VGCPs using the range-Doppler model. The word “independent” implies that the virtual checkpoints are never used in the RPC fitting, i.e., they are located in different positions respective to the VGCPs. The accuracy of the fitted RPCs for the TerraSAR-X data in each study area is listed in Table \ref{tab.RPCresult}. Analysis was performed based on the residuals of the rows and columns, given by the differences $ row_{RPC}-row_{DR}$ and $ col_{RPC}-col_{DR}$, i.e., the differences between image coordinates computed by RPCs and range-Doppler. The analysis results confirm that the RPCs can model the range-Doppler geometry for TerraSAR-X data to within a millimeter, and can thus well be used in the 3D reconstruction process.    

\begin{table}[h]
	\centering \footnotesize
	\caption{Accuracy (standard deviation: STD) of RPCs fitted on SAR sensor model (units: m)}
	\label{tab.RPCresult}
	\begin{tabular}{l |cc|cc}
		
		\multirow{2}{*}{\textbf{Area}}&  \multicolumn{2}{c|}{\textbf{Virtual GCPs}}& \multicolumn{2}{c}{\textbf{Check points}} \\
		&\textbf{row} &\textbf{column}&\textbf{row} &\textbf{column} \\\hline
        \toprule
		Munich 	 		& 0.00026& 0.00114 &0.00025 &0.00031 \\
		Berlin 	        & 0.00024& 0.00027 &0.00026 &0.00118 \\
	\end{tabular} 	
\end{table}

\subsection{Validity of the Epipolarity Constraint}\label{subsec.dis-epi}
A general model that proves the epipolarity constraint for SAR-optical image pairs was described in Section \ref{Epi}. It was also concluded that epipolar curves are usually not straight. Experimentally, the epipolarity constraint for SAR-optical image pairs can also be modeled based on RPCs. In this paper, we evaluate the epipolarity constraint for TerraSAR-X and WorldView-2 image pairs acquired over the two study areas.

\subsubsection{Existence of Epipolar Curves}\label{subsubsec.dis-epi1}
\begin{figure*} 
	\centering
	\subfloat[TerraSAR-X: Munich]{%
		\includegraphics[width=0.49\linewidth]{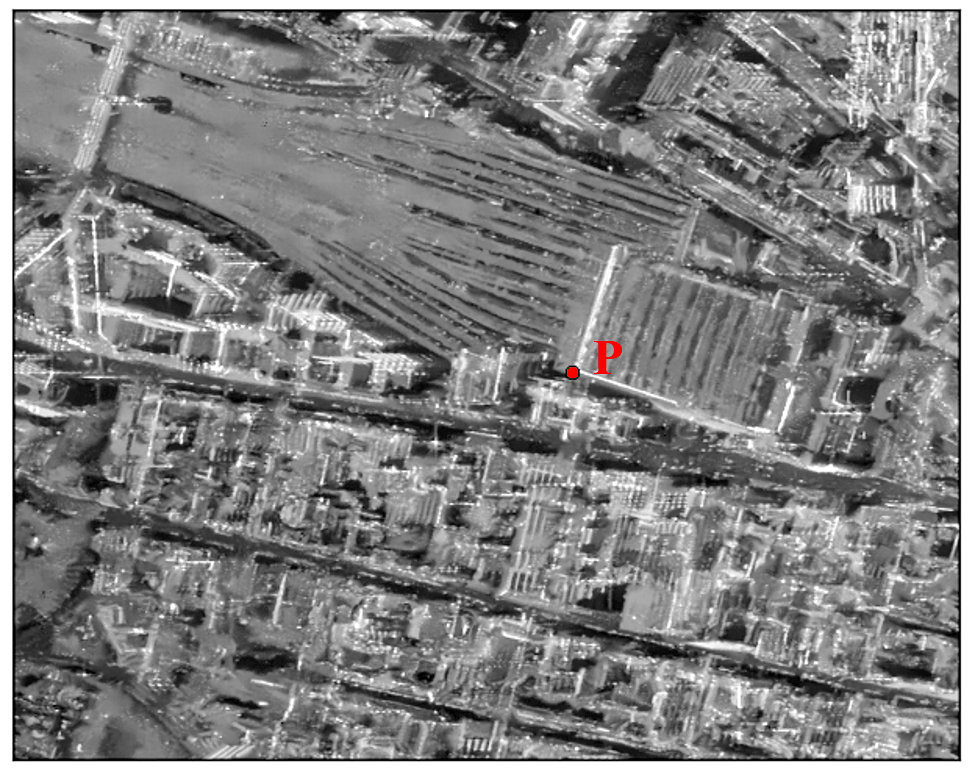}}
	\label{fig.epi1}
	\subfloat[WorldView-2: Munich]{%
		\includegraphics[width=0.49\linewidth]{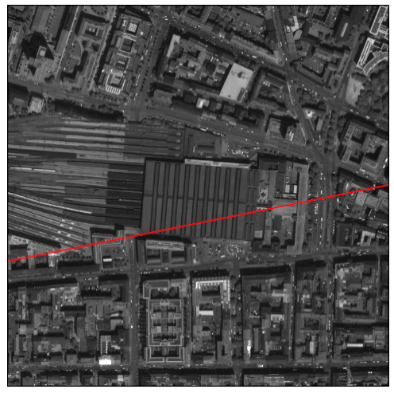}}
	\label{fig.epi2}
	
	\subfloat[WorldView-2: Munich]{%
		\includegraphics[width=0.9\linewidth]{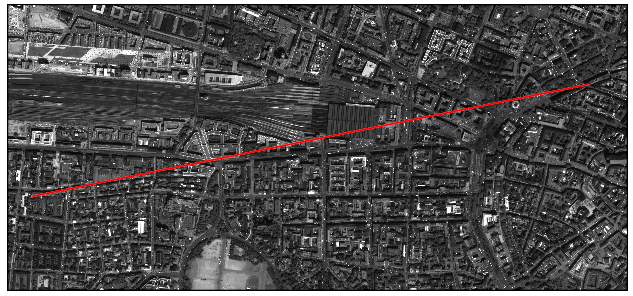}}
	\label{fig.epi3}
	
	\caption{Epipolar curves for the WorldView-2 image (of Munich) given by changing the heights of point p (located at the corner of the Munich central train station in the TerraSAR-X scene) for all possible height values in the image scene. The epipolar curves look like straight, but are not actually straight.}
	\label{fig.epiM} 
\end{figure*}

\begin{figure*} 
	\centering
	
	\subfloat[TerraSAR-X: Berlin]{%
		\includegraphics[width=0.49\linewidth]{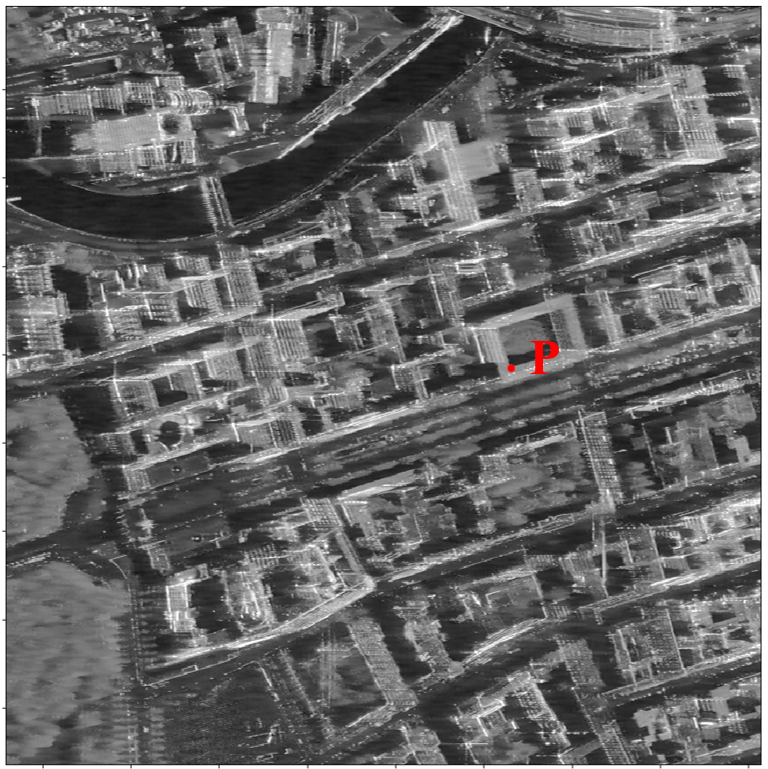}}
	\label{fig.epi4}
	\subfloat[WorldView-2: Berlin]{%
		\includegraphics[width=0.49\linewidth]{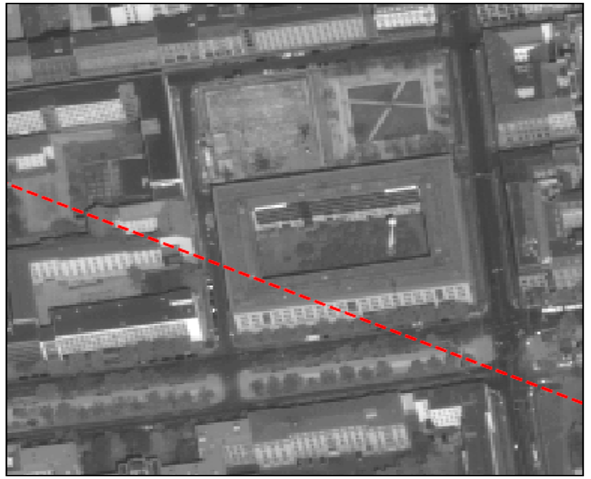}}
	\label{fig.epi6}
	
	\subfloat[WorldView-2: Berlin]{%
		\includegraphics[width=0.8\linewidth]{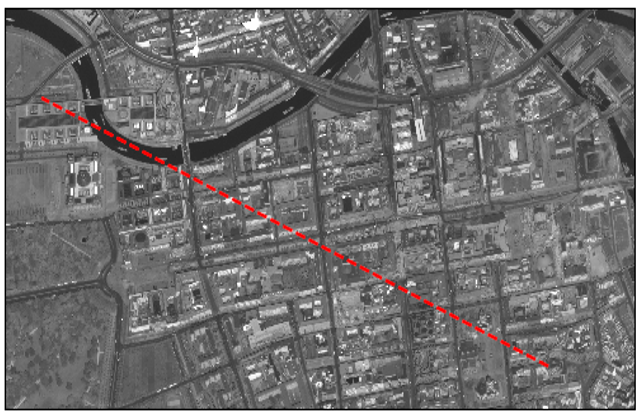}}
	\label{fig.epi5}
	
	\caption{Epipolar curves for the WorldView-2 image (of Berlin) given by changing the heights of point p (a distinct corner of a building in the Berlin TerraSAR-X scene) for all possible height values in the image scene. The epipolar curves look like straight, but are not actually straight.}
	\label{fig.epiB} 
\end{figure*}

We analyzed the validity of the derived SAR-optical epipolarity constraint for an exemplary point located at the corner of the Munich central train station building ($ p $). This point was projected to the terrain space by changing the heights in specific steps, e.g., 10 $ m $, starting from the lowest possible height and proceeding to the highest possible height in the scene (for this experiment we used the interval [0 $ m $, 1200 $ m $]).  The output will be an ensemble of points with different heights, such as depicted in Fig. \ref{fig.epiM}(c). All these points were then back-projected to the WorldView-2 image space using RPCs. The corresponding epipolar curve for all possible heights in the study area is constructed by connecting the image points obtained in this way, as shown in Fig. \ref{fig.epiM}(c). Although the epipolar curve appears to be straight, more analysis is required to determine whether this is the case. By expanding the image, it can be seen in Fig. \ref{fig.epiM}(b) that the epipolar curve nearly passes through the conjugate point of $ p $ in the WorldView-2 image. Similarly, another experiment was carried out using the Berlin dataset. The epipolar curve was constructed for an exemplary point located on the corner of a building and for all possible heights in the scene. Figure \ref{fig.epiB}(d) displays the position of the selected point on the SAR image as well as the corresponding epipolar curve in the WorldView-2 imagery (Figs. \ref{fig.epiB}(e) and \ref{fig.epiB}(f)).             

\subsubsection{Straightness of Epipolar Curves}\label{subsubsec.dis-epi2}
\begin{figure} 
	\centering
	\subfloat[Munich]{%
		\includegraphics[width=0.49\linewidth]{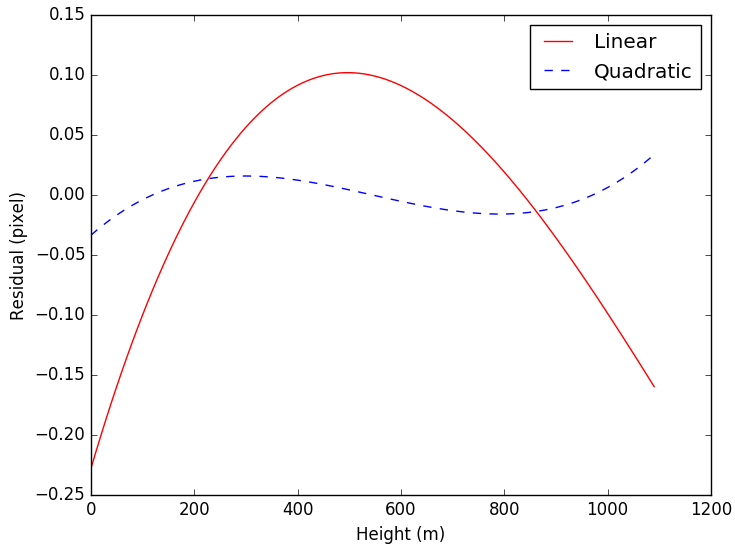}}
	\label{fig.epi-cur1}
	\subfloat[Berlin]{%
		\includegraphics[width=0.49\linewidth]{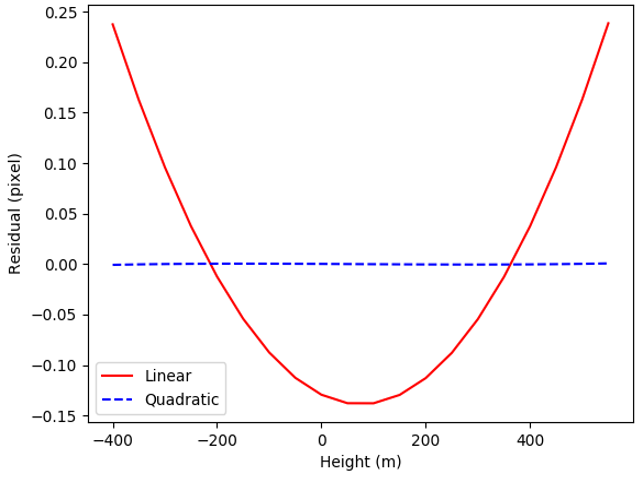}}
	\label{fig.epi-cur2}
	\caption{Linear and quadratic polynomials fitted on the epipolar curves in the WorldView-2 images}
	\label{fig.epi-cur} 
\end{figure}

To clarify the straightness of the epipolar curve constructed for point $p$, linear and quadratic polynomials were fitted to the image points of the epipolar curve. Figures  \ref{fig.epi-cur}(a) and  \ref{fig.epi-cur}(b)  represent the least-squares residuals with respect to the point heights for the epipolar curves created in both study subsets. The residuals of the linear fit for the epipolar curve established for the Munich WorldView-2 image range from -0.25 to 0.1 pixels (i.e. meters), whereas the residuals of the quadratic fit are close to zero. 

Similar results were given by fitting linear and quadratic polynomials to the image points of the epipolar curve established in the WorldView-2 image of Berlin. Figure \ref{fig.epi-cur}(b) clearly shows that the residuals of the epipolar points fitted to the quadratic model are zero, whereas those of the linear fit vary between -0.15 $ m $ and 0.25 $ m $. Both analyses illustrate that the constructed epipolar curves are not straight. However, their curvatures do not exceed more than one pixel over the whole possible range of heights in these scenes.      

\subsubsection{Conjugacy of Epipolar Curves}\label{subsubsec.dis-epi3}
\begin{figure*}[ht!]
	\begin{center}
		\includegraphics[width=1\textwidth]{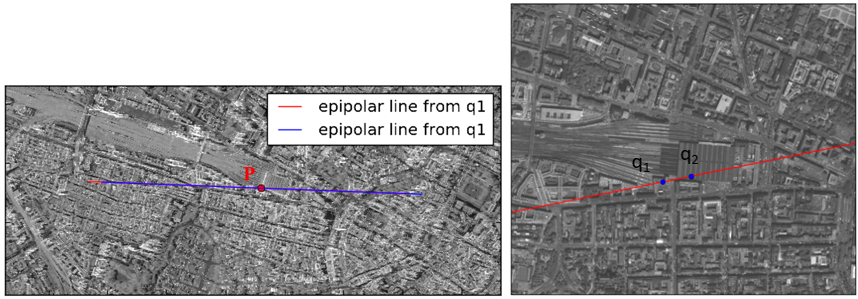}
		\caption{Corresponding epipolar curves in the Munich TerraSAR-X image (left) derived from two points, $q_{1}$ and $q_{2}$ on the epipolar curve of the Munich WorldView-2 image (right).}
		\label{fig.epi-resid}
	\end{center}
\end{figure*}

\begin{figure} 
	\centering
	\subfloat[Munich]{%
		\includegraphics[width=0.49\linewidth]{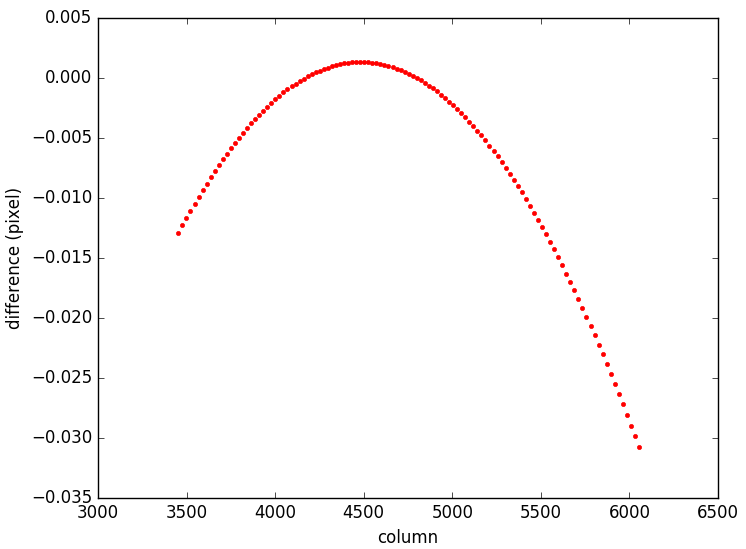}}
	\label{fig.epi-pair3}	
	\subfloat[Berlin]{%
		\includegraphics[width=0.49\linewidth]{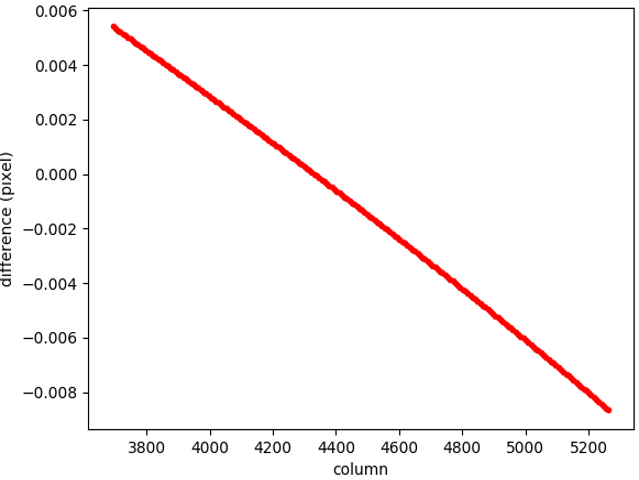}}
	\label{fig.epi-pair3_1}
	\caption{Difference of two corresponding epipolar curves over the column direction. The maximum difference between the two epipolar curves is less than one pixel.}
	\label{fig.epi-pair} 
\end{figure}

\begin{figure} 
	\centering
	\subfloat[Munich]{%
		\includegraphics[width=0.49\linewidth]{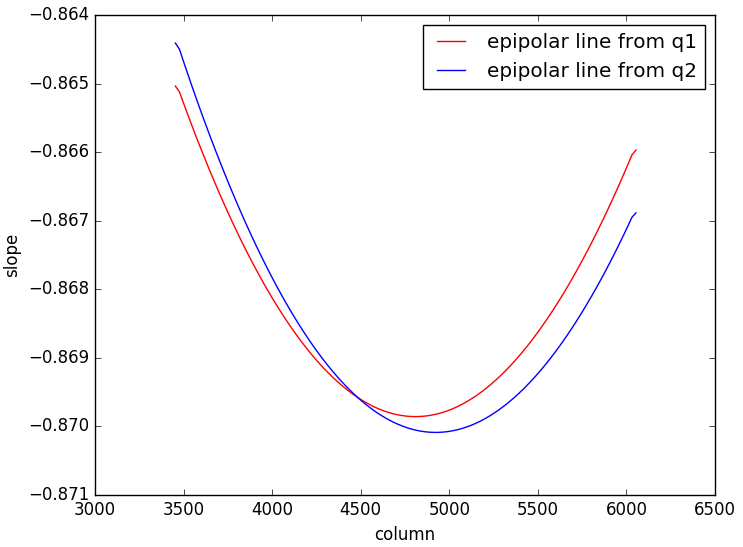}}
	\label{fig.epi-pair4}	
	\subfloat[Berlin]{%
		\includegraphics[width=0.49\linewidth]{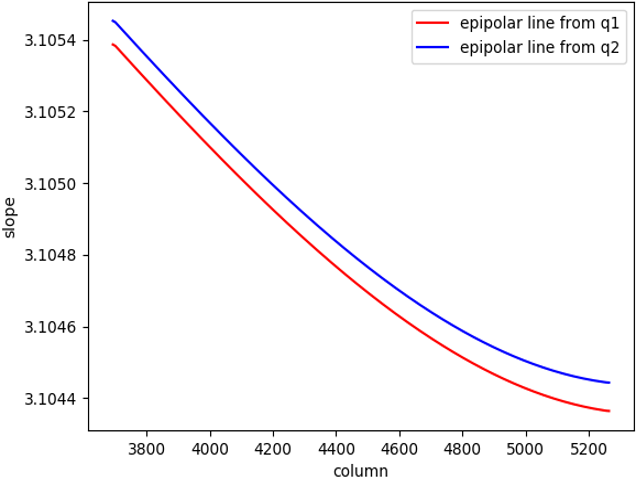}}
	\label{fig.epi-pair4_1}
	\caption{Gradients of two epipolar curves constructed from q1 and q2 in TerraSAR-X.}
	\label{fig.epi-pairn} 
\end{figure}

To investigate the conjugacy of the SAR-optical epipolar curves, two distinct points ($q_{1}$,$q_{2}$) were selected from the epipolar curve in the WorldView-2 image. From each of these points, the corresponding epipolar curves were constructed in the TerraSAR-X image for all possible heights as in the experiments before.
Figure \ref{fig.epi-resid} displays the corresponding epipolar curves in TerraSAR-X given by $q_{1}$ and $q_{2}$ located in the WorldView-2 image. The epipolar curve appears to pass through point $p$ located in the SAR image. Further analysis clarifies that the differences in the column direction between the two epipolar curves passing through point $p$ are less than one pixel, allowing the matching of the two epipolar curves (Fig. \ref{fig.epi-pair}(a)). In addition, Fig. \ref{fig.epi-pairn}(a) shows that the gradients of the two epipolar curves change at each point, as illustrated by the column index, whereas the maximum difference is less than 0.001, i.e., 0.1\%. This indicates that the epipolar curves can be assumed to be parallel. The gradient changes at each point also confirm that the epipolar curves in TerraSAR-X are not perfectly straight. 

In a similar manner, the conjugacy of the epipolar curves was evaluated for the Berlin dataset. Figure \ref{fig.epi-pair}(b) shows that the difference between the epipolar curves is less than one pixel, so these lines can be paired. Similarly, the maximum difference between the slopes of the epipolar curves is less than 0.2\%,  which confirms the possibility of epipolar curve conjugacy (Fig. \ref{fig.epi-pairn}(b)).      

From the above investigations and discussions, it is clear that the epipolarity constraint can be established for SAR-optical image pairs such as those from TerraSAR-X and WorldView-2 data. As expected, the epipolar curves are not perfectly straight and there are tiny differences between the epipolar curve in one image produced from points on the epipolar curve in the other image. However, analyses show that the epipolar curves can be approximated as straight lines without sacrificing too much, and that they can be paired together well. This means that the epipolarity constraint can be used to ease the subsequent stereogrammetric matching process. 

\subsection{Use of Block Adjustment}\label{subsec.dis-badj}
As discussed in Section \ref{subsec.dis-epi} and mathematically proved in Section \ref{Epi}, the epipolarity constraint can be established for a SAR-optical image pair. However, the positions of epipolar curves in the optical image can be placed in a more accurate position by exploiting the high geolocalization accuracy of the SAR image through a multi-sensor block adjustment. The experiments described in Section \ref{subsec.dis-epi} demonstrate that, for the case of WorldView-2 imagery, the curvature of the epipolar curves does not exceed one pixel, and using only two bias terms as shifts in the column and row directions suffice to modify the position of the epipolar curves.

\begin{figure} 
	\centering
	\subfloat[Munich]{%
		\includegraphics[width=0.49\linewidth]{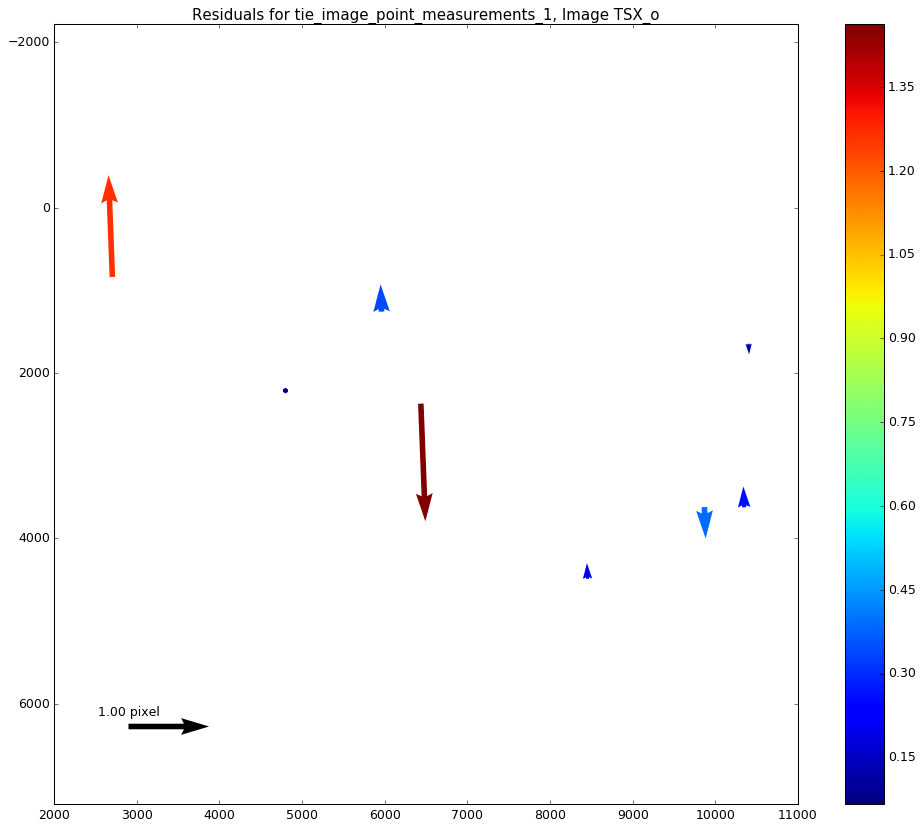}}
	\label{fig.BA_munich}
	\subfloat[Berlin]{%
		\includegraphics[width=0.49\linewidth]{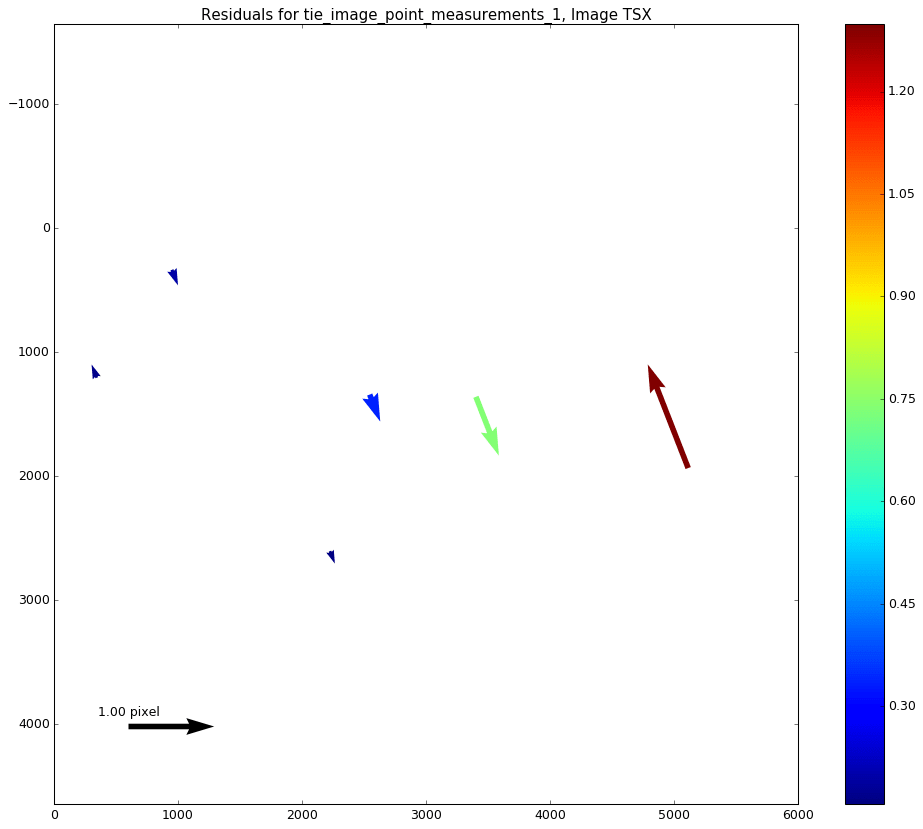}}
	\label{fig.BA_berlin}
	\caption{Residuals of tie points after full multi-sensor block adjustment in the TerraSAR-X image space.}
	\label{fig.BAresid} 
\end{figure}

\begin{figure*} 
	\centering
	\subfloat[Munich]{%
		\includegraphics[width=0.8\linewidth]{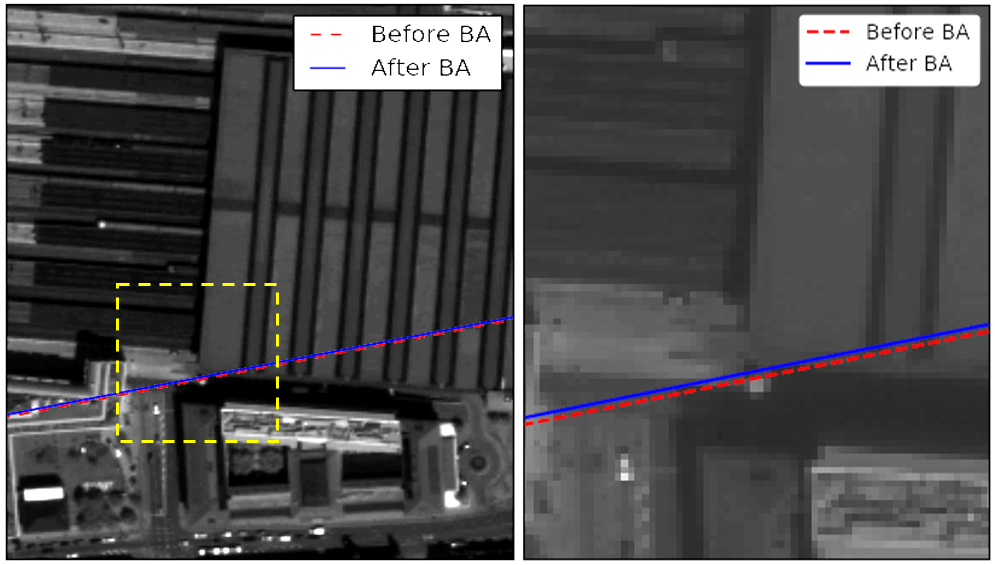}}
	\label{fig.BA_epi_muinch}
	
	\subfloat[Berlin]{%
		\includegraphics[width=0.9\linewidth]{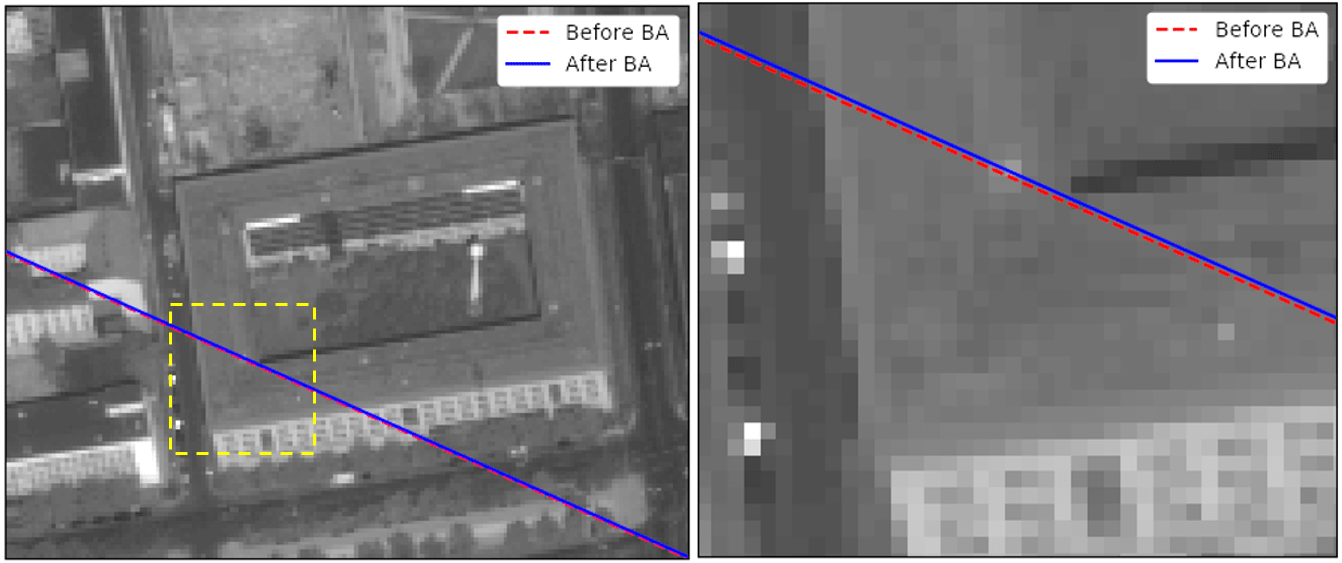}}
	\label{fig.BA_epi_berlin}
	
	\caption{Displacement of epipolar curves after block adjustment by RPCs. Left images show the epipolar curve positions before and after the bundle adjustment and right images display the selected patch (identified by dashed yellow rectangles) in an enlarged image.}
	\label{fig.BA_epi} 
\end{figure*}

Implementing block adjustment using RPCs requires some conjugate points to be assigned as common tie points between the target (WorldView-2) and the reference (TerraSAR-X) images. Theoretically, just one tie point would be sufficient to estimate the bias in the least-squares adjustment based on equation (\ref{eq.BA}) (two unknowns and two equations), but using more redundancy and incorporating more tie points allows for more accurate estimations of the bias parameter. For this experiment, eight and six tie points were selected to match the WorldView-2 images to the TerraSAR-X images in the Munich and Berlin study areas, respectively. The block adjustment equations were then established  as described in Section \ref{blockAdj}. During the iterative least-squares adjustment, tie points with residuals exceeding a threshold were removed from the full adjustment process.  Figures \ref{fig.BAresid}(a) and \ref{fig.BAresid}(b) show the residuals of the full multi-sensor block adjustment for each tie point. The results demonstrate that the residuals of most points are less than one pixel in both experiments, which indicates a successful implementation of SAR-optical block adjustment.  
Table \ref{BA-result} presents the bias of the row and column components resulting from the block adjustment of WorldView-2 and TerraSAR-X image pairs for both study areas. As the SAR image was selected as the reference to which the optical imagery was aligned, the bias components for the reference SAR imagery are zero. The quality of block adjustment has been evaluated by calculating the positional errors of tie points under projection and reprojection from the SAR image to terrain and from terrain to the optical image, and vice versa. Statistical metrics such as the standard deviation (STD), median absolute deviation (MAD), and minimum (Min)/maximum (Max) errors were calculated. The results illustrate that the major bias component is in the row direction in both study areas and that the epipolar curves will mostly shift in this direction after block adjustment.
Figures \ref{fig.BA_epi}(a) and \ref{fig.BA_epi}(b) display the locations of the epipolar curves before and after adjustment. By enlarging the images, it is possible to confirm that the displacement of the epipolar curves is minimal, yet noticeable.

\begin{table}[h]
	\centering \footnotesize
	\caption{Block adjustment results (units: m)}
	\label{BA-result}
	\begin{tabular}{l c|cccccc|c}
		\textbf{Area} &\textbf{Sensor} &  \multicolumn{2}{c} {\thead {\textbf{Bias} \\ \textbf{Coefficients}}}& \textbf{STD} & \textbf{MAD}& \textbf{Min}&\textbf{Max}& \textbf{\thead{ No. of \\ Tie Points}}  \\
        \toprule
		\multirow{2}{*}{Munich} 
		&WorldView-2 &-2.47 &-0.53	&0.50 & 0.14& 0.07& 1.46 & 8\\
		&TerraSAR-X	 &0 &0	&0.50 & 0.14& 0.07& 1.46 & \\
        \midrule
		\multirow{2}{*}{Berlin}
		&WorldView-2 &-0.73&0.28& 0.51 & 0.13& 0.19& 1.59 & 6\\
        &TerraSAR-X	 &0 &0	&0.42 & 0.11& 0.16& 1.30 & \\
		\hline		
	\end{tabular}
\end{table}

To verify the success of SAR-optical block adjustment, we require highly accurate GCPs, which are not available for the study areas. However, we evaluated the accuracy of block adjustment in sub-scenes of the two study areas with the assistance of available LiDAR point clouds. First, we manually found some matched points that had been measured in the SAR and optical sub-scenes, similar to the tie point selection step. Next, the measured points located in the optical imagery were projected to the terrain using the corresponding reverse rational polynomial functions  ($ f_{o}^{\prime}(c,r, h) $ and $ g_{o}^{\prime}(c,r, h) $). To ensure exact back-projection, the height $h$ of each point was extracted from the available high-resolution LiDAR point clouds of the target sub-scenes. To overcome the noise in the LiDAR data, we considered neighboring points around the selected measured point, and the final height of the target point was selected based on the mode of the heights in the considered neighborhood. The resulting ground points ($ f_{o}^{\prime}(c,r, h) $, $ g_{o}^{\prime}(c,r, h) $, $ h $) were then back-projected to the SAR scene using the forward RPCs fitted to the SAR imagery. Finally, comparing the image coordinates of the measured points on the SAR imagery (from manual matching) with their coordinates derived by projection from the optical to the SAR imagery using RPCs and LiDAR data provides an evaluation of the SAR-optical block adjustment performance. The residuals can be calculated as:
 
\begin{equation}\label{eq.BAverify}
\begin{tabular}{c} 
$ dc = f_{s}(f_{o}^{\prime}(c,r, h),  g_{o}^{\prime}(c,r, h), h)-c_{s}^{m} $ \\
$ dr = g_{s}(f_{o}^{\prime}(c,r, h),  g_{o}^{\prime}(c,r, h), h)-r_{s}^{m} $
\end{tabular}
\end{equation} 
where $ (dc, dr) $ is the column and row difference between the measured point $ (c_{s}^{m}, r_{s}^{m}) $ located on the SAR imagery and the corresponding coordinates given by the projection from the optical to the SAR imagery using RPCs.

Table \ref{BA-verification} presents some statistical analysis on residuals calculated according to equation (\ref{eq.BAverify}) for two states: using the original WorldView-2 RPCs for the projections and using the WorldView-2 RPCs modified with respect to the TerraSAR-X SAR imagery. The results demonstrate the successful implementation of RPC-based multi-sensor block adjustment for SAR-optical image pairs. This means that the existing bias in the RPCs of optical imagery such as WorldView-2 can be modified according to high-resolution SAR imagery such as TerraSAR-X to improve the absolute geolocalization accuracy of the optical imagery, and consequently the modification of epipolar curves in stereo cases.

\begin{table}[h]
	\centering \footnotesize
	\caption{Residuals (unit: m) of the control points for block adjustment validation. \emph{Original} indicates the residuals before the adjustment, \emph{modified} those after adjustment.}
	\label{BA-verification}
	\begin{tabular}{l c|ccc|c}
		\textbf{Area}   & \textbf{RPCs}& \textbf{Mean} & \textbf{Max}&\textbf{RMSE}& \thead{\textbf{No. of} \\ \textbf{Points}}  \\\hline
        \toprule
		\multirow{2}{*}{Munich} 
		&Original  &1.301 &2.359& 1.364	&31 \\
		&Modified  &0.666 &2.142& 0.923& \\
        \midrule
		\multirow{2}{*}{Berlin}
		&Original  &0.775&1.736& 0.866& 32\\
		&Modified  &0.600&1.600& 0.752&\\

		\hline		
	\end{tabular}
\end{table}      

\subsection{Dense Matching Results}\label{subsec.result} 
\begin{figure*} 
	\centering
	\subfloat[Munich]{%
		\includegraphics[width=0.45\linewidth, height= 0.6\linewidth]{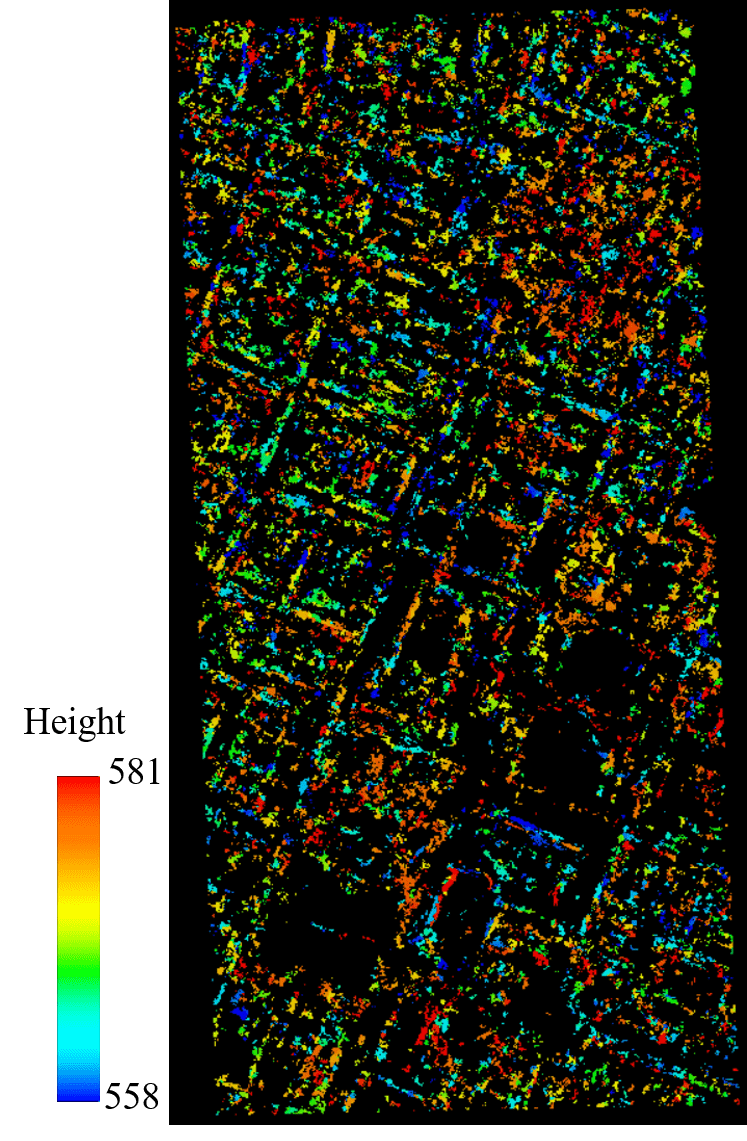}}
	\label{fig:munich3d}
	\subfloat[Berlin]{%
		\includegraphics[width=0.4\linewidth, height= 0.6\linewidth]{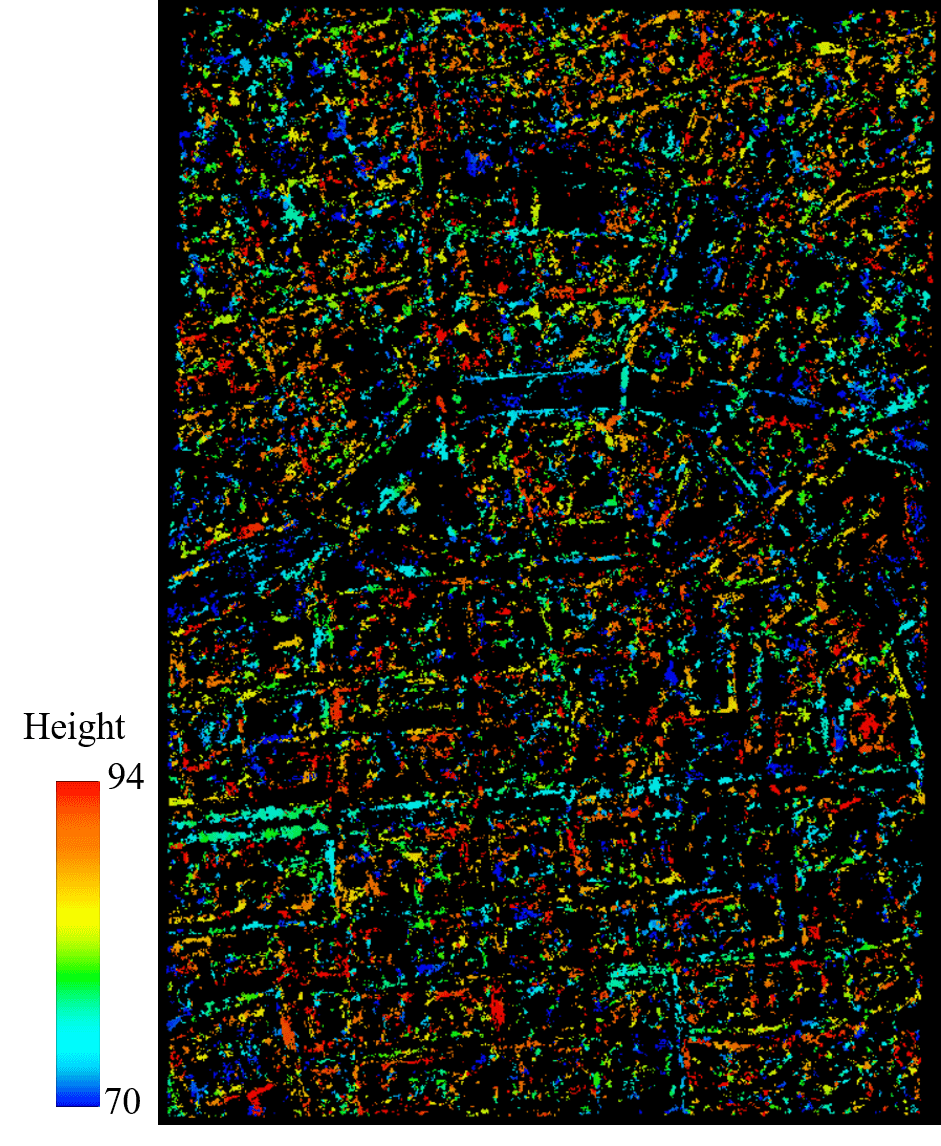}}
	\label{fig:berlin3d}
	\caption{Point clouds reconstructed from Munich and Berlin sub-scenes}
	\label{fig:3dpc} 
\end{figure*}

The output of dense matching by SGM is a disparity map that is calculated in the frame of the reference sensor geometry. This disparity map should be transferred from the reference sensor geometry to a terrestrial reference coordinate system such as UTM. The difference of SAR and optical observation geometries and the lack of jointly visible scene parts means that stereogrammetric 3D reconstruction leads to sparse rather than dense point clouds over urban areas. Figures \ref{fig:3dpc}(a) and \ref{fig:3dpc}(b) display the reconstructed point clouds from SAR-optical sub-scenes of Munich and Berlin.

The accuracy of these sparse point clouds was compared to that of reference LiDAR point clouds with densities of 6 and 6.5 points per square meter acquired by airborne sensors over Munich and Berlin, respectively. 

Different approaches can be used to assess the accuracy of point clouds. The simplest way is to calculate the Euclidean distance to the nearest-neighbor of each target point in the reference point cloud \cite{Muja2009}. This strategy, however, should only be used when both point clouds are very dense. We therefore used another approach, which is based on fitting a plane to the $k$ (here is 6) nearest neighbors of each target point in the reference point cloud \cite{Mitra2004}. The perpendicular distance from the target point to this plane is then the measured reconstruction error. To speed up the process of point cloud evaluation, an octree data structure is used for the binary partitioning of both reconstructed and reference point clouds \cite{Schnabel2006}. The measured distances between both point clouds are decomposed into three components that represent the accuracy of the reconstructed points along the $X$, $Y$, and $Z$ directions. 
Table \ref{tab.accuracyPC} summarizes the mean, STD, and Root Mean Square Error (RMSE) of the distances along the different axes.

\begin{table}[h]
	\centering
	\footnotesize
	\caption{Accuracy assessment of reconstructed point clouds with respect to LiDAR reference}
	\label{tab.accuracyPC}
	\begin{tabular}{l|ccc|ccc|ccc}
		\multirow{2}{*}{\textbf{Area}} & \multicolumn{3}{c}{\textbf{Mean (m)}} &  \multicolumn{3}{c}{\textbf{STD (m)}} &  \multicolumn{3}{c}{\textbf{RMSE (m)}}\\
        
			& \textbf{X} & \textbf{Y} & \textbf{Z} & \textbf{X} & \textbf{Y} & \textbf{Z} & \textbf{X} & \textbf{Y} & \textbf{Z}\\\hline
        \toprule
		Munich	       	&-0.003	&0.025	&0.080 &1.285 &1.350 &2.652 &1.285 &1.351 &2.653 \\
    	\midrule
		Berlin 	       	&0.000	&-0.041	&0.273&1.566 &1.692 &3.091 &1.566 &1.693 &3.103 \\
	\end{tabular} 
\end{table}

\begin{figure}[ht!]
\centering
\subfloat[before outlier removal]{
\includegraphics[width=0.35\linewidth]{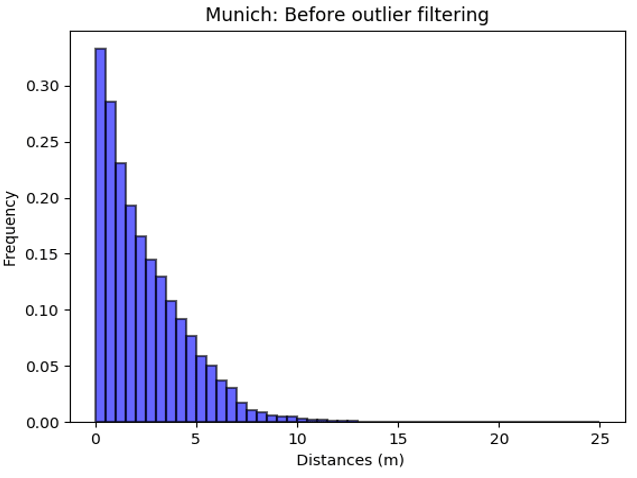}}
\subfloat[after outlier removal]{
\includegraphics[width=0.35\linewidth]{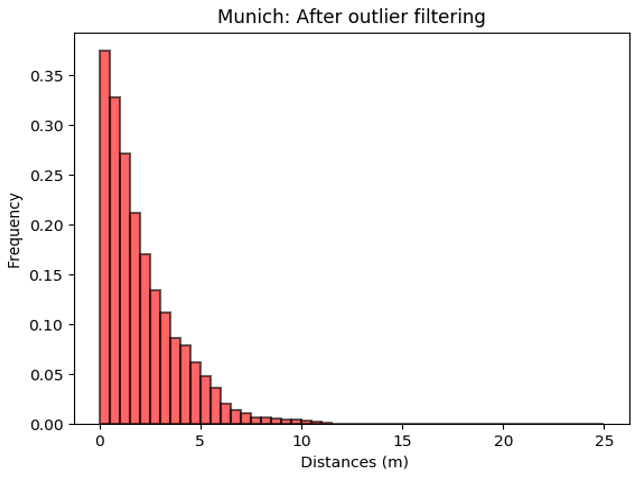}}
\caption{Euclidian distances between reconstructed points and reference planes for Munich.}\label{fig:distancesMunich}
\end{figure}

\begin{figure}[ht!]
\centering
\subfloat[before outlier removal]{
\includegraphics[width=0.35\linewidth]{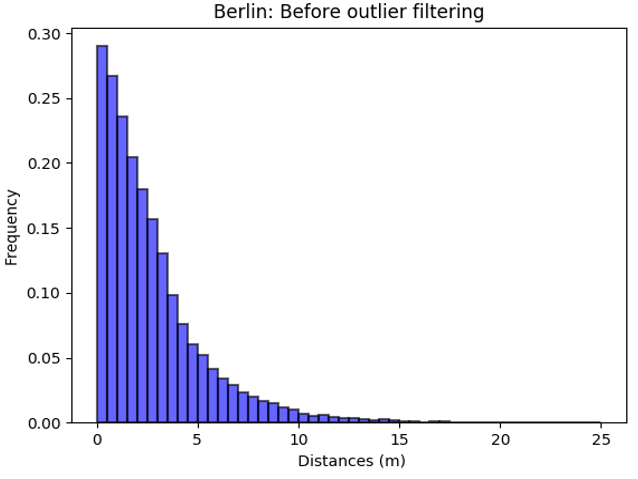}}
\subfloat[after outlier removal]{
\includegraphics[width=0.35\linewidth]{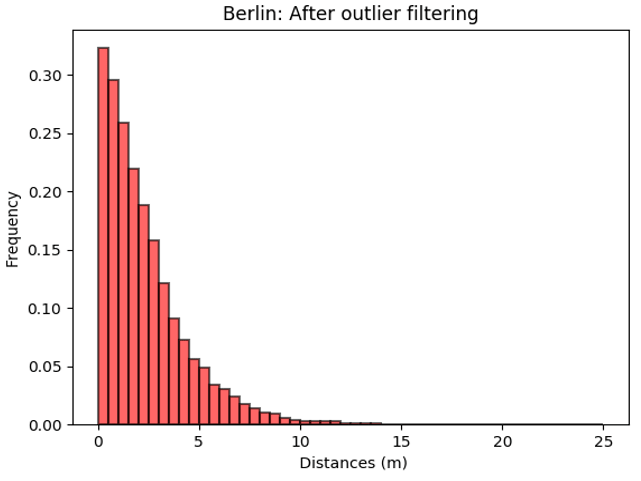}}
\caption{Euclidian distances between reconstructed points and reference planes for Berlin.}\label{fig:distancesBerlin}
\end{figure}

In addition, histograms of the Euclidean distances between reconstructed points and $k$-nearest neighbors-based reference planes are depicted in Figs.~\ref{fig:distancesMunich} and \ref{fig:distancesBerlin}, while the corresponding metrics are summarized in Tab.~\ref{tab.accuracy_filteredPC}. In order to also provide an outlier-free accuracy assessment, we additionally show results corresponding to point clouds that were cleaned by removing points deviating from the SRTM model by more than 5m.

\begin{table}[h]
	\centering\footnotesize
	\caption{Accuracy assessment of point clouds after SRTM-based outlier removal}
	\label{tab.accuracy_filteredPC}
	\begin{tabular}{l l |ccc|c}
		\textbf{Area}&\textbf{Point Cloud} & \textbf{25\%-quantile}& \textbf{50\%-quantile} & \textbf{75\%-quantile}& \textbf{Mean (m)}\\
		\toprule
		\multirow{3}{*}{Munich}	&original &0.77 & 1.89 & 3.58 & 2.44\\
					&filtered &0.67& 1.56 & 3.04 &2.12 \\
					&\textit{SRTM} &\textit{0.73} &\textit{1.64} &\textit{3.25} &\textit{2.21}\\
		\midrule
		\multirow{2}{*}{Berlin} &original &0.89& 2.01 & 3.67 &2.75\\
					&filtered &0.79 & 1.76 & 3.22 & 2.35\\
					&\textit{SRTM} &\textit{0.86}&\textit{1.93}&\textit{3.63}&\textit{2.65}\\
	\end{tabular} 
\end{table}

\begin{figure}[ht!]
	\begin{center}
		\includegraphics[width=0.6\linewidth]{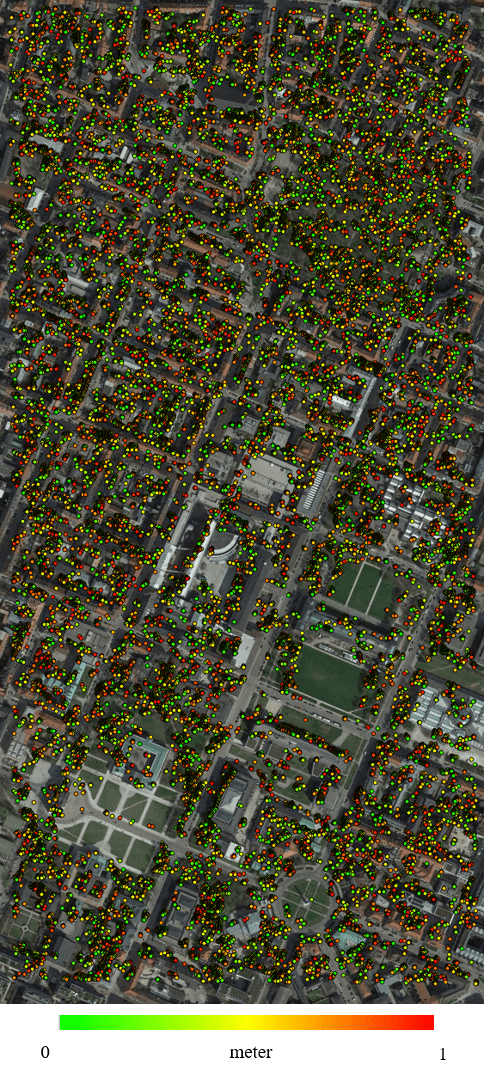}
		\caption{Position of points with subpixel accuracy (1 $ m $) achieved by dense matching
			over Munich city}
		\label{fig.munich-residual}
	\end{center}
\end{figure}

\begin{figure}[ht!]
	\begin{center}
		\includegraphics[width=0.8\linewidth]{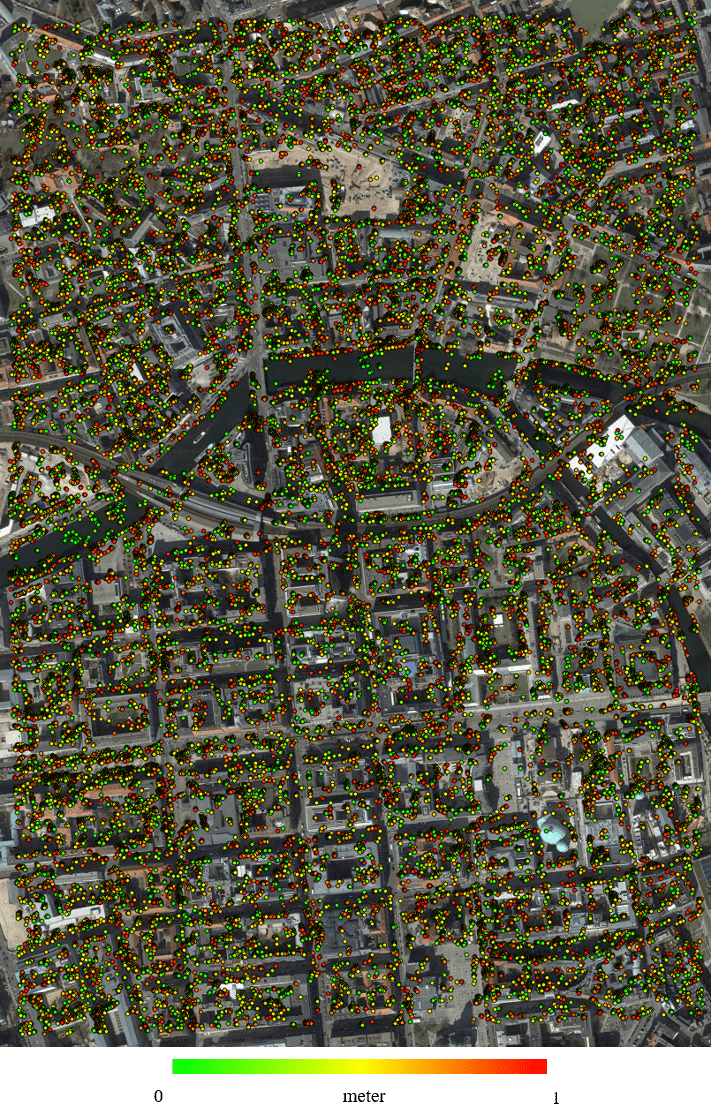}
		\caption{Position of points with subpixel accuracy (1 $ m $) achieved by dense matching
			over Berlin city}
		\label{fig.berlin-residual}
	\end{center}
\end{figure}

Finally, Figs. \ref{fig.munich-residual} and \ref{fig.berlin-residual} display high-accuracy points (i.e. those with a Euclidean distance of less than 1m to the reference) achieved by SGM dense matching for TerraSAR-X- WorldView-2 image pairs in Munich and Berlin, respectively.


\section{Discussion}\label{sec.dis} 
\subsection{Feasibility of SAR-Optical Stereogrammetry Workflow}\label{sec.feasibility}
The results described in the previous section demonstrate the potential of the proposed SAR-optical stereogrammetry framework. Our analyses show that all primary steps involved in SAR-optical stereogrammetry, such as RPC fitting, epipolar-curve generation, and multi-sensor block adjustment, can be successfully implemented for VHR SAR-optical image pairs. In addition to the mathematical proof of the existence of an epipolarity constraint for arbitrary SAR-optical image pairs in Section \ref{Epi}, the experimental results have illustrated the validity of establishing an epipolarity constraint by showing that SAR-optical epipolar curves are approximately straight. Using RPCs for both sensor types paves the way for the implementation of stereogrammetry. As a result, estimating the RPCs for SAR imagery is a prerequisite for SAR-optical stereogrammetry. The RPCs delivered with optical imagery must be improved with respect to the SAR sensor geometry using RPC-based multi-sensor block adjustment. The block adjustment aligns pairs of SAR and optical images and improves the absolute geopositioning accuracy of the optical imagery. This ensures that the epipolar curves pass through the correct positions of conjugate points. Applying a dense matching algorithm such as SGM then produces a disparity map.

\subsection{Potential and Limitations of SAR-Optical Stereogrammetry}\label{sec.limit}  
\begin{figure}[ht!]
	\begin{center}
		\includegraphics[width=0.5\linewidth]{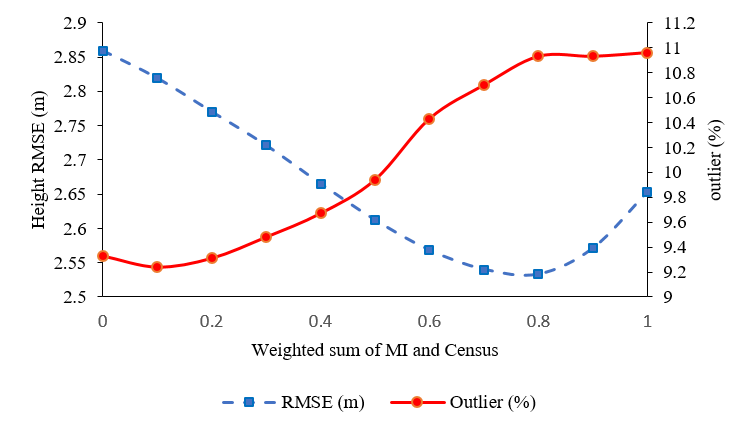}
		\caption{Performance of MI, Census, and weighted sum of both measures as cost functions in SGM.}
		\label{fig.MIcensus}
	\end{center}
\end{figure}

As discussed in Section \ref{subsec.result}, the dense matching of TerraSAR-X/WorldView-2 imagery produces a sparse point cloud over each of the urban study areas. However, the resulting point clouds are affected by a significant amount of noise because of the difficult radiometric and geometric relationships between the SAR and the optical images. Hence, the SGM algorithm struggles to find the exact conjugate points. On the one hand, this is related to the similarity measures employed in this prototypical study. The influence of similarity measures on the height accuracy of the Munich point cloud is shown in Fig. \ref{fig.MIcensus}. 

The RMSE of the estimated heights decreases when Census and MI are combined and used as a weighted sum cost function, although the number of outliers increases. Identifying the optimum weighting to balance the percentage of outliers against the height accuracy is impractical, because the output disparity maps are rather sparse; a visualization would not be helpful for this task. In stereogrammetric 3D reconstruction using optical image pairs, visualizing the disparity map enables the weight value to be tuned so as to preserve the edges and sharpness of building footprints, whereas in the SAR-optical case, there is no perfectly dense disparity map. Using Census alone produces points with higher accuracy than the MI-only results, but a higher percentage of outliers. In general, a similarity measure specifically designed for SAR-optical matching is required.  

On the other hand, the reconstruction suffers from the fact that the SGM search strategy is designed for relatively simple isotropic geometric distortions and was not adapted to the peculiarities of SAR-optical matching yet. Therefore, the differences in the imaging geometries of SAR and optical sensors in terms of their off-nadir and horizontal viewing angles can further decrease the matching accuracy. For example it is known from previous research that the optimal geometrical condition for SAR-optical stereogrammetry would be an image pair acquired with similar viewing geometries \cite{Qiu2018218}. This, however, would make the geometrically induced dissimilarities in the images even larger and render the matching more complicated. If both sensors were at the same position, and thus would share the same viewing angle, the intersection geometry would be perfect \cite{Qiu2018218}. However, due to the different imaging geometries, elevated objects would appear to collapse away from the sensor in the optical image, while they would appear to collapse towards the sensor in the SAR image. Thus, the choice of a good stereo geometry will always need to be a trade-of between image similarity and favorable intersection angle in the SAR-optical case.

Last, but not least, many points cannot be sensed by a nadir-looking optical sensor but are well observed by a side-looking SAR sensor, such as points located on building facades. As has been shown before, the joint visibility between SAR and optical VHR images of urban scenes can be as low as about 50\% \cite{Hughes2018}. In the present study, the situation was most complicated in the Berlin case, because of differences in both the horizontal viewing directions and the off-nadir angles. The horizontal viewing direction of the WorldView-2 sensor was approximately north-south, whereas that of TerraSAR-X was east-west. This affected the visibility of common points between the two images during 3D reconstruction negatively. Consequently, most of the reconstructed points are located on the flat areas or outlines of buildings that are observed by both sensors (see Figs. \ref{fig.munich-residual} and \ref{fig.berlin-residual}).  

Finally, some differences in the image pairs may be caused by the interval between the acquisition times of the WorldView-2 and TerraSAR-X data (5 and 3 years for Munich and Berlin, respectively). This can cause the matching process to fail in problematic areas, thus affecting the quality and density of the disparity maps.

In spite of the differences in sensor geometries, acquisition times, and illumination conditions between the two SAR-optical image pairs, the quantitative analysis demonstrated in Section~\ref{subsec.result} shows that 25\% of all points are reconstructed with clear sub-pixel accuracy, while the median accuracy lies at about 1.5 to 2m.
The experiments also show that the results can be further improved by filtering outliers from the reconstructed point clouds. In this study, we employed the globally available SRTM DEM as prior knowledge for outlier removal. As Tab. \ref{tab.accuracy_filteredPC} shows, discarding points with a height difference to SRTM greater than 5m improves the results significantly.

Of course, this simple filtering strategy will probably also remove some accurate points that just deviate a lot from the SRTM DEM (e.g. newly built skyscrapers). In conclusion, a more sophisticated algorithm should be developed for removing noise and outliers from derived point clouds in the future. Nevertheless it can be confirmed that the SAR-optical stereo results have the potential to provide both higher accuracy and higher point density than the SRTM data, making SAR-optical stereogrammetry another possible means for 3D reconstruction in remote sensing.

\section{Conclusion}\label{sec.conclusion}

In this study, we investigated the possibility of stereogrammetric 3D reconstruction from VHR SAR-optical image pairs by developing a full 3D reconstruction framework based on the classic photogrammetric workflow. First, we analyzed all prerequisites for this task. The main requirement for SAR-optical stereogrammetry is to establish an epipolarity constraint to reduce the search space of the matching process. We mathematically proved that the epipolarity constraint can be established for SAR-optical image pairs. Furthermore, experimental analysis demonstrated that the epipolarity constraint can be employed for SAR-optical image pairs such as those from TerraSAR-X/WorldView-2, and showed that the epipolar curves are sufficiently straight. Because of the limited accuracy of the RPCs delivered with optical data, the relative orientation between both images can be improved with respect to the more accurate SAR orientation parameters using multi-sensor block adjustment. This shifts the epipolar curves toward their correct positions. An SGM-based dense matching algorithm was implemented using the MI and Census similarity measures, as well as their weighted sum. The outputs were sparse point clouds with a median accuracy of about 1.5 to 2m and the 25\%-quantile of best points well in the sub-pixel accuracy domain. Finally, SRTM data were used to remove outliers from the point clouds. This improved the accuracy of the point clouds further. Overall, this study has demonstrated that a 3D reconstruction framework can be designed and implemented for SAR-optical image pairs over urban areas. Future work will have to focus on the development of similarity metrics specific to the multi-sensor matching problem, and on an adaption of the semi-global search strategy that accounts for the anisotropic geometric distortions between SAR and optical images.


\section*{Acknowledgments}
The authors want to thank everyone, who has provided test data
for this research: European Space Imaging for the WorldView-2 image, DLR for the TerraSAR-X images, the Bavarian Surveying Administration for
the LiDAR reference data of Munich, and Land Berlin (EU EFRE project) for the LiDAR reference data of Berlin.

This work is jointly supported by the German Research Foundation (DFG) under grant SCHM 3322/1-1, the Helmholtz Association under the framework of the
Young Investigators Group SiPEO (VH-NG-1018), and the European Research Council (ERC) under the European Union’s Horizon 2020 research and innovation programme (grant agreement No. ERC-2016-StG-714087, Acronym:
So2Sat).



\section*{References}
\bibliographystyle{model1-num-names}
\bibliography{sample.bib}

\begin{thebibliography}{49}
\expandafter\ifx\csname natexlab\endcsname\relax\def\natexlab#1{#1}\fi
\providecommand{\bibinfo}[2]{#2}
\ifx\xfnm\relax \def\xfnm[#1]{\unskip,\space#1}\fi
\bibitem[{Schmitt and Zhu(2016)}]{Schmitt2016a}
\bibinfo{author}{M.~Schmitt}, \bibinfo{author}{X.~X. Zhu},
\newblock \bibinfo{title}{Data fusion and remote sensing: An ever-growing
  relationship},
\newblock \bibinfo{journal}{IEEE Geoscience and Remote Sensing Magazine}
  \bibinfo{volume}{4} (\bibinfo{year}{2016}) \bibinfo{pages}{6--23}.
\bibitem[{Bloom et~al.(1988)Bloom, Fielding, and Fu}]{Bloom1988}
\bibinfo{author}{A.~L. Bloom}, \bibinfo{author}{E.~J. Fielding},
  \bibinfo{author}{X.-Y. Fu},
\newblock \bibinfo{title}{A demonstration of stereophotogrammetry with combined
  {SIR-B} and {Landsat TM} images},
\newblock \bibinfo{journal}{International Journal of Remote Sensing}
  \bibinfo{volume}{9} (\bibinfo{year}{1988}) \bibinfo{pages}{1023--1038}.
\bibitem[{{J. Raggam, A. Almer, and D. Strobl}(1994)}]{raggam1994}
\bibinfo{author}{{J. Raggam, A. Almer, and D. Strobl}},
\newblock \bibinfo{title}{A combination of {SAR} and optical line scanner
  imagery for stereoscopic extraction of 3{D} data},
\newblock \bibinfo{journal}{ISPRS Journal of Photogrammetry and Remote Sensing}
  \bibinfo{volume}{49} (\bibinfo{year}{1994}) \bibinfo{pages}{11--21}.
\bibitem[{Wegner et~al.(2014)Wegner, Ziehn, and Soergel}]{6690226}
\bibinfo{author}{J.~D. Wegner}, \bibinfo{author}{J.~R. Ziehn},
  \bibinfo{author}{U.~Soergel},
\newblock \bibinfo{title}{Combining high-resolution optical and {InSAR}
  features for height estimation of buildings with flat roofs},
\newblock \bibinfo{journal}{IEEE Transactions on Geoscience and Remote Sensing}
  \bibinfo{volume}{52} (\bibinfo{year}{2014}) \bibinfo{pages}{5840--5854}.
\bibitem[{Morgan et~al.(2004)Morgan, El-Sheimy, and
  Saalfeld}]{Morgan2004EpipolarRO}
\bibinfo{author}{M.~F. Morgan}, \bibinfo{author}{N.~El-Sheimy},
  \bibinfo{author}{A.~Saalfeld}, \bibinfo{title}{Epipolar resampling of linear
  array scanner scenes}, \bibinfo{type}{{PhD} dissertation}, University of
  Calgary, Department of Geomatics Engineering, \bibinfo{year}{2004}.
\bibitem[{Scharstein et~al.(2001)Scharstein, Szeliski, and Zabih}]{988771}
\bibinfo{author}{D.~Scharstein}, \bibinfo{author}{R.~Szeliski},
  \bibinfo{author}{R.~Zabih},
\newblock \bibinfo{title}{A taxonomy and evaluation of dense two-frame stereo
  correspondence algorithms},
\newblock in: \bibinfo{booktitle}{Proceedings IEEE Workshop on Stereo and
  Multi-Baseline Vision (SMBV 2001)}, pp. \bibinfo{pages}{131--140}.
\bibitem[{Eineder et~al.(2011)Eineder, Minet, Steigenberger, Cong, and
  Fritz}]{5570983}
\bibinfo{author}{M.~Eineder}, \bibinfo{author}{C.~Minet},
  \bibinfo{author}{P.~Steigenberger}, \bibinfo{author}{X.~Cong},
  \bibinfo{author}{T.~Fritz},
\newblock \bibinfo{title}{Imaging geodesy--toward centimeter-level ranging
  accuracy with {TerraSAR-X}},
\newblock \bibinfo{journal}{IEEE Transactions on Geoscience and Remote Sensing}
  \bibinfo{volume}{49} (\bibinfo{year}{2011}) \bibinfo{pages}{661--671}.
\bibitem[{DigitalGlobe(2018)}]{digitalglobe}
\bibinfo{author}{DigitalGlobe}, \bibinfo{title}{Accuracy of {WorldView}
  products},
  \bibinfo{howpublished}{\url{https://dg-cms-uploads-production.s3.amazonaws.com/uploads/document/file/38/DG_ACCURACY_WP_V3.pdf}},
  \bibinfo{year}{2018}. \bibinfo{note}{(Accessed 03.18)}.
\bibitem[{Bagheri et~al.(2018)Bagheri, Schmitt, d'Angelo, and
  Zhu}]{Bagheri2018}
\bibinfo{author}{H.~Bagheri}, \bibinfo{author}{M.~Schmitt},
  \bibinfo{author}{P.~d'Angelo}, \bibinfo{author}{X.~X. Zhu},
\newblock \bibinfo{title}{Exploring the applicability of semi-global matching
  for {SAR}-optical stereogrammetry of urban scenes},
\newblock \bibinfo{journal}{ISPRS - International Archives of the
  Photogrammetry, Remote Sensing and Spatial Information Sciences}
  \bibinfo{volume}{42(2)} (\bibinfo{year}{2018}) \bibinfo{pages}{43--48}.
\bibitem[{Oh et~al.(2010)Oh, Lee, Toth, Grejner-Brzezinska, and
  Lee}]{oh2010piecewise}
\bibinfo{author}{J.~Oh}, \bibinfo{author}{W.~H. Lee}, \bibinfo{author}{C.~K.
  Toth}, \bibinfo{author}{D.~A. Grejner-Brzezinska}, \bibinfo{author}{C.~Lee},
\newblock \bibinfo{title}{A piecewise approach to epipolar resampling of
  pushbroom satellite images based on {RPC}},
\newblock \bibinfo{journal}{Photogrammetric Engineering \& Remote Sensing}
  \bibinfo{volume}{76} (\bibinfo{year}{2010}) \bibinfo{pages}{1353--1363}.
\bibitem[{Grodecki and Dial(2003)}]{grodecki2003block}
\bibinfo{author}{J.~Grodecki}, \bibinfo{author}{G.~Dial},
\newblock \bibinfo{title}{Block adjustment of high-resolution satellite images
  described by rational polynomials},
\newblock \bibinfo{journal}{Photogrammetric Engineering \& Remote Sensing}
  \bibinfo{volume}{69} (\bibinfo{year}{2003}) \bibinfo{pages}{59--68}.
\bibitem[{Fraser et~al.(2006)Fraser, Dial, and Grodecki}]{FRASER2006182}
\bibinfo{author}{C.~Fraser}, \bibinfo{author}{G.~Dial},
  \bibinfo{author}{J.~Grodecki},
\newblock \bibinfo{title}{Sensor orientation via {RPC}s},
\newblock \bibinfo{journal}{ISPRS Journal of Photogrammetry and Remote Sensing}
  \bibinfo{volume}{60} (\bibinfo{year}{2006}) \bibinfo{pages}{182 -- 194}.
\bibitem[{Li et~al.(2007)Li, Zhou, Niu, and Di}]{li2007integration}
\bibinfo{author}{R.~Li}, \bibinfo{author}{F.~Zhou}, \bibinfo{author}{X.~Niu},
  \bibinfo{author}{K.~Di},
\newblock \bibinfo{title}{Integration of {Ikonos} and {QuickBird} imagery for
  geopositioning accuracy analysis},
\newblock \bibinfo{journal}{Photogrammetric Engineering and Remote Sensing}
  \bibinfo{volume}{73} (\bibinfo{year}{2007}) \bibinfo{pages}{1067}.
\bibitem[{Tao and Hu(2002)}]{tao20023d}
\bibinfo{author}{C.~V. Tao}, \bibinfo{author}{Y.~Hu},
\newblock \bibinfo{title}{3{D} reconstruction methods based on the rational
  function model},
\newblock \bibinfo{journal}{Photogrammetric Engineering \& Remote Sensing}
  \bibinfo{volume}{68} (\bibinfo{year}{2002}) \bibinfo{pages}{705--714}.
\bibitem[{Tao et~al.(2004)Tao, Hu, and
  Jiang}]{doi:10.1080/01431160310001618392}
\bibinfo{author}{C.~V. Tao}, \bibinfo{author}{Y.~Hu},
  \bibinfo{author}{W.~Jiang},
\newblock \bibinfo{title}{Photogrammetric exploitation of {Ikonos} imagery for
  mapping applications},
\newblock \bibinfo{journal}{International Journal of Remote Sensing}
  \bibinfo{volume}{25} (\bibinfo{year}{2004}) \bibinfo{pages}{2833--2853}.
\bibitem[{Toutin(2006)}]{toutin2006comparison}
\bibinfo{author}{T.~Toutin},
\newblock \bibinfo{title}{Comparison of 3{D} physical and empirical models for
  generating {DSMs} from stereo {HR} images},
\newblock \bibinfo{journal}{Photogrammetric Engineering \& Remote Sensing}
  \bibinfo{volume}{72} (\bibinfo{year}{2006}) \bibinfo{pages}{597--604}.
\bibitem[{Tao and Hu(2001)}]{doi:10.1080/07038992.2001.10854900}
\bibinfo{author}{C.~Tao}, \bibinfo{author}{Y.~Hu},
\newblock \bibinfo{title}{Use of the rational function model for image
  rectification},
\newblock \bibinfo{journal}{Canadian Journal of Remote Sensing}
  \bibinfo{volume}{27} (\bibinfo{year}{2001}) \bibinfo{pages}{593--602}.
\bibitem[{Grodecki et~al.(2004)Grodecki, Dial, and Lutes}]{Grodecki2004}
\bibinfo{author}{J.~Grodecki}, \bibinfo{author}{G.~Dial},
  \bibinfo{author}{J.~Lutes},
\newblock \bibinfo{title}{Mathematical model for 3{D} feature extraction from
  multiple satellite images described by {RPCs}},
\newblock in: \bibinfo{booktitle}{ASPRS Annual Conference Proceedings, Denver,
  Colorado}, \bibinfo{year}{2004}.
\bibitem[{Tao and Hu(2001)}]{tao2001comprehensive}
\bibinfo{author}{C.~V. Tao}, \bibinfo{author}{Y.~Hu},
\newblock \bibinfo{title}{A comprehensive study of the rational function model
  for photogrammetric processing},
\newblock \bibinfo{journal}{Photogrammetric Engineering and Remote Sensing}
  \bibinfo{volume}{67} (\bibinfo{year}{2001}) \bibinfo{pages}{1347--1358}.
\bibitem[{Tikhonov and Arsenin(1977)}]{tikhonov1977methods}
\bibinfo{author}{A.~Tikhonov}, \bibinfo{author}{V.~Y. Arsenin},
  \bibinfo{title}{Methods for solving ill-posed problems},
  \bibinfo{publisher}{John Wiley and Sons, Inc}, \bibinfo{year}{1977}.
\bibitem[{Zhang et~al.(2011)Zhang, He, Balz, Wei, and Liao}]{ZHANG2011133}
\bibinfo{author}{L.~Zhang}, \bibinfo{author}{X.~He}, \bibinfo{author}{T.~Balz},
  \bibinfo{author}{X.~Wei}, \bibinfo{author}{M.~Liao},
\newblock \bibinfo{title}{Rational function modeling for spaceborne {SAR}
  datasets},
\newblock \bibinfo{journal}{ISPRS Journal of Photogrammetry and Remote Sensing}
  \bibinfo{volume}{66} (\bibinfo{year}{2011}) \bibinfo{pages}{133 -- 145}.
\bibitem[{Cho et~al.(1993)Cho, Schenk, and Madani}]{cho1993resampling}
\bibinfo{author}{W.~Cho}, \bibinfo{author}{T.~Schenk},
  \bibinfo{author}{M.~Madani},
\newblock \bibinfo{title}{Resampling digital imagery to epipolar geometry},
\newblock \bibinfo{journal}{International Archives of Photogrammetry and Remote
  Sensing} \bibinfo{volume}{29} (\bibinfo{year}{1993})
  \bibinfo{pages}{404--404}.
\bibitem[{Gupta and Hartley(1997)}]{615446}
\bibinfo{author}{R.~Gupta}, \bibinfo{author}{R.~I. Hartley},
\newblock \bibinfo{title}{Linear pushbroom cameras},
\newblock \bibinfo{journal}{IEEE Transactions on Pattern Analysis and Machine
  Intelligence} \bibinfo{volume}{19} (\bibinfo{year}{1997})
  \bibinfo{pages}{963--975}.
\bibitem[{Kim(2000)}]{kim2000study}
\bibinfo{author}{T.~Kim},
\newblock \bibinfo{title}{A study on the epipolarity of linear pushbroom
  images},
\newblock \bibinfo{journal}{Photogrammetric Engineering \& Remote Sensing}
  \bibinfo{volume}{66} (\bibinfo{year}{2000}) \bibinfo{pages}{961--966}.
\bibitem[{Orun(1994)}]{orun1994modified}
\bibinfo{author}{A.~B. Orun},
\newblock \bibinfo{title}{A modified bundle adjustment software for {SPOT}
  imagery and photography: tradeoff},
\newblock \bibinfo{journal}{Photogrammetric Engineering \& Remote Sensing}
  \bibinfo{volume}{60} (\bibinfo{year}{1994}) \bibinfo{pages}{1431--1437}.
\bibitem[{Morgan et~al.(2004)Morgan, Kim, Jeong, and
  Habib}]{morgan2004epipolar}
\bibinfo{author}{M.~Morgan}, \bibinfo{author}{K.~Kim},
  \bibinfo{author}{S.~Jeong}, \bibinfo{author}{A.~Habib},
\newblock \bibinfo{title}{Epipolar geometry of linear array scanners moving
  with constant velocity and constant attitude},
\newblock \bibinfo{journal}{ISPRS - International Archives of the
  Photogrammetry, Remote Sensing and Spatial Information Sciences}
  \bibinfo{volume}{35} (\bibinfo{year}{2004}) \bibinfo{pages}{508--513}.
\bibitem[{Curlander(1982)}]{4157311}
\bibinfo{author}{J.~C. Curlander},
\newblock \bibinfo{title}{Location of spaceborne {SAR} imagery},
\newblock \bibinfo{journal}{IEEE Transactions on Geoscience and Remote Sensing}
  \bibinfo{volume}{22} (\bibinfo{year}{1982}) \bibinfo{pages}{106--112}.
\bibitem[{Gutjahr et~al.(2014)Gutjahr, Perko, Raggam, and
  Schardt}]{Gutjahr2014TheEC}
\bibinfo{author}{K.~Gutjahr}, \bibinfo{author}{R.~Perko},
  \bibinfo{author}{J.~Raggam}, \bibinfo{author}{M.~Schardt},
\newblock \bibinfo{title}{The epipolarity constraint in stereo-radargrammetric
  {DEM} generation},
\newblock \bibinfo{journal}{IEEE Transactions on Geoscience and Remote Sensing}
  \bibinfo{volume}{52} (\bibinfo{year}{2014}) \bibinfo{pages}{5014--5022}.
\bibitem[{Li and Zhang(2013)}]{6509489}
\bibinfo{author}{D.~Li}, \bibinfo{author}{Y.~Zhang},
\newblock \bibinfo{title}{A rigorous sar epipolar geometry modeling and
  application to 3{D} target reconstruction},
\newblock \bibinfo{journal}{IEEE Journal of Selected Topics in Applied Earth
  Observations and Remote Sensing} \bibinfo{volume}{6} (\bibinfo{year}{2013})
  \bibinfo{pages}{2316--2323}.
\bibitem[{Hartley and Zisserman(2004)}]{Hartley2004}
\bibinfo{author}{R.~I. Hartley}, \bibinfo{author}{A.~Zisserman},
  \bibinfo{title}{Multiple View Geometry in Computer Vision},
  \bibinfo{publisher}{Cambridge University Press, ISBN: 0521540518},
  \bibinfo{edition}{second} edition, \bibinfo{year}{2004}.
\bibitem[{Kratky(1989)}]{kratky1988rigorous}
\bibinfo{author}{V.~Kratky},
\newblock \bibinfo{title}{Rigorous photogrammetric processing of {SPOT} images
  at {CCM Canada}},
\newblock \bibinfo{journal}{ISPRS Journal of Photogrammetry and Remote Sensing}
  \bibinfo{volume}{44} (\bibinfo{year}{1989}) \bibinfo{pages}{53 -- 71}.
\bibitem[{Curlander and McDonough(1991)}]{curlander1991synthetic}
\bibinfo{author}{J.~C. Curlander}, \bibinfo{author}{R.~N. McDonough},
  \bibinfo{title}{Synthetic aperture radar}, volume \bibinfo{volume}{396},
  \bibinfo{publisher}{John Wiley \& Sons New York, NY, USA},
  \bibinfo{year}{1991}.
\bibitem[{d'Angelo and Reinartz(2012)}]{dlr78910}
\bibinfo{author}{P.~d'Angelo}, \bibinfo{author}{P.~Reinartz},
\newblock \bibinfo{title}{{DSM} based orientation of large stereo satellite
  image blocks},
\newblock \bibinfo{journal}{ISPRS - International Archives of the
  Photogrammetry, Remote Sensing and Spatial Information Sciences}
  \bibinfo{volume}{39(B1)} (\bibinfo{year}{2012}) \bibinfo{pages}{209--214}.
\bibitem[{Tong et~al.(2010)Tong, Liu, and Weng}]{TONG2010218}
\bibinfo{author}{X.~Tong}, \bibinfo{author}{S.~Liu}, \bibinfo{author}{Q.~Weng},
\newblock \bibinfo{title}{Bias-corrected rational polynomial coefficients for
  high accuracy geo-positioning of {QuickBird} stereo imagery},
\newblock \bibinfo{journal}{ISPRS Journal of Photogrammetry and Remote Sensing}
  \bibinfo{volume}{65} (\bibinfo{year}{2010}) \bibinfo{pages}{218 -- 226}.
\bibitem[{Fraser and Hanley(2005)}]{fraser2005bias}
\bibinfo{author}{C.~S. Fraser}, \bibinfo{author}{H.~B. Hanley},
\newblock \bibinfo{title}{Bias-compensated {RPCs} for sensor orientation of
  high-resolution satellite imagery},
\newblock \bibinfo{journal}{Photogrammetric Engineering \& Remote Sensing}
  \bibinfo{volume}{71} (\bibinfo{year}{2005}) \bibinfo{pages}{909--915}.
\bibitem[{Suri and Reinartz(2010)}]{5340570}
\bibinfo{author}{S.~Suri}, \bibinfo{author}{P.~Reinartz},
\newblock \bibinfo{title}{Mutual-information-based registration of {TerraSAR-X}
  and {Ikonos} imagery in urban areas},
\newblock \bibinfo{journal}{IEEE Transactions on Geoscience and Remote Sensing}
  \bibinfo{volume}{48} (\bibinfo{year}{2010}) \bibinfo{pages}{939--949}.
\bibitem[{Perko et~al.(2011)Perko, Raggam, Gutjahr, and Schardt}]{6049732}
\bibinfo{author}{R.~Perko}, \bibinfo{author}{H.~Raggam},
  \bibinfo{author}{K.~Gutjahr}, \bibinfo{author}{M.~Schardt},
\newblock \bibinfo{title}{Using worldwide available {TerraSAR-X} data to
  calibrate the geo-location accuracy of optical sensors},
\newblock in: \bibinfo{booktitle}{2011 IEEE International Geoscience and Remote
  Sensing Symposium}, pp. \bibinfo{pages}{2551--2554}.
\bibitem[{Merkle et~al.(2017)Merkle, Luo, Auer, M{\"u}ller, and
  Urtasun}]{Merkle2017}
\bibinfo{author}{N.~Merkle}, \bibinfo{author}{W.~Luo},
  \bibinfo{author}{S.~Auer}, \bibinfo{author}{R.~M{\"u}ller},
  \bibinfo{author}{R.~Urtasun},
\newblock \bibinfo{title}{Exploiting deep matching and {SAR} data for the
  geo-localization accuracy improvement of optical satellite images},
\newblock \bibinfo{journal}{Remote Sensing} \bibinfo{volume}{9}
  (\bibinfo{year}{2017}) \bibinfo{pages}{586}.
\bibitem[{Brown et~al.(2003)Brown, Burschka, and Hager}]{Brown2003}
\bibinfo{author}{M.~Z. Brown}, \bibinfo{author}{D.~Burschka},
  \bibinfo{author}{G.~D. Hager},
\newblock \bibinfo{title}{Advances in computational stereo},
\newblock \bibinfo{journal}{IEEE Transactions on Pattern Analysis and Machine
  Intelligence} \bibinfo{volume}{25} (\bibinfo{year}{2003})
  \bibinfo{pages}{993--1008}.
\bibitem[{Hirschm{\"u}ller(2008)}]{Hirschmuller2008}
\bibinfo{author}{H.~Hirschm{\"u}ller},
\newblock \bibinfo{title}{Stereo processing by semiglobal matching and mutual
  information},
\newblock \bibinfo{journal}{IEEE Transactions on Pattern Analysis and Machine
  Intelligence} \bibinfo{volume}{30} (\bibinfo{year}{2008})
  \bibinfo{pages}{328--341}.
\bibitem[{Hassaballah et~al.(2016)Hassaballah, Abdelmgeid, and
  Alshazly}]{Hassaballah2016}
\bibinfo{author}{M.~Hassaballah}, \bibinfo{author}{A.~A. Abdelmgeid},
  \bibinfo{author}{H.~A. Alshazly},
\newblock \bibinfo{title}{Image features detection, description and matching},
\newblock in: \bibinfo{editor}{A.~I. Awad}, \bibinfo{editor}{M.~Hassaballah}
  (Eds.), \bibinfo{booktitle}{Image Feature Detectors and Descriptors :
  Foundations and Applications}, \bibinfo{publisher}{Springer International
  Publishing}, \bibinfo{year}{2016}, pp. \bibinfo{pages}{11--45}.
\bibitem[{Zhu et~al.(2011)Zhu, d'Angelo, and
  Butenuth}]{10.1007/978-3-642-24393-6_14}
\bibinfo{author}{K.~Zhu}, \bibinfo{author}{P.~d'Angelo},
  \bibinfo{author}{M.~Butenuth},
\newblock \bibinfo{title}{A performance study on different stereo matching
  costs using airborne image sequences and satellite images},
\newblock in: \bibinfo{editor}{U.~Stilla}, \bibinfo{editor}{F.~Rottensteiner},
  \bibinfo{editor}{H.~Mayer}, \bibinfo{editor}{B.~Jutzi},
  \bibinfo{editor}{M.~Butenuth} (Eds.), \bibinfo{booktitle}{Photogrammetric
  Image Analysis}, \bibinfo{publisher}{Springer Berlin Heidelberg},
  \bibinfo{address}{Berlin, Heidelberg}, \bibinfo{year}{2011}, pp.
  \bibinfo{pages}{159--170}.
\bibitem[{USGS(2000)}]{USGS2000}
\bibinfo{author}{USGS}, \bibinfo{title}{{S}huttle {R}adar {T}opography
  {M}ission ({SRTM}) void filled},
  \bibinfo{howpublished}{\url{https://lta.cr.usgs.gov/SRTMVF}},
  \bibinfo{year}{2000}. \bibinfo{note}{(accessed 09.17)}.
\bibitem[{Egnal and Wildes(2002)}]{Egnal2002}
\bibinfo{author}{G.~Egnal}, \bibinfo{author}{R.~P. Wildes},
\newblock \bibinfo{title}{Detecting binocular half-occlusions: empirical
  comparisons of five approaches},
\newblock \bibinfo{journal}{IEEE Transactions on Pattern Analysis and Machine
  Intelligence} \bibinfo{volume}{24} (\bibinfo{year}{2002})
  \bibinfo{pages}{1127--1133}.
\bibitem[{Muja and Lowe(2009)}]{Muja2009}
\bibinfo{author}{M.~Muja}, \bibinfo{author}{D.~G. Lowe},
\newblock \bibinfo{title}{Fast approximate nearest neighbors with automatic
  algorithm configuration},
\newblock in: \bibinfo{booktitle}{In VISAPP International Conference on
  Computer Vision Theory and Applications}, \bibinfo{year}{2009}, pp.
  \bibinfo{pages}{331--340}.
\bibitem[{Mitra et~al.(2004)Mitra, Nguyen, and Guibas}]{Mitra2004}
\bibinfo{author}{N.~J. Mitra}, \bibinfo{author}{A.~Nguyen},
  \bibinfo{author}{L.~Guibas},
\newblock \bibinfo{title}{Estimating surface normals in noisy point cloud
  data},
\newblock \bibinfo{journal}{International Journal of Computational Geometry \&
  Applications} \bibinfo{volume}{14} (\bibinfo{year}{2004})
  \bibinfo{pages}{261--276}.
\bibitem[{Schnabel and Klein(2006)}]{Schnabel2006}
\bibinfo{author}{R.~Schnabel}, \bibinfo{author}{R.~Klein},
\newblock \bibinfo{title}{Octree-based point-cloud compression},
\newblock in: \bibinfo{booktitle}{Proceedings of the 3rd Eurographics / IEEE
  VGTC Conference on Point-Based Graphics}, SPBG'06,
  \bibinfo{publisher}{Eurographics Association},
  \bibinfo{address}{Aire-la-Ville, Switzerland, Switzerland},
  \bibinfo{year}{2006}, pp. \bibinfo{pages}{111--121}.
\bibitem[{Qiu et~al.(2018)Qiu, Schmitt, and Zhu}]{Qiu2018218}
\bibinfo{author}{C.~Qiu}, \bibinfo{author}{M.~Schmitt}, \bibinfo{author}{X.~X.
  Zhu},
\newblock \bibinfo{title}{Towards automatic {SAR}-optical stereogrammetry over
  urban areas using very high resolution imagery},
\newblock \bibinfo{journal}{ISPRS Journal of Photogrammetry and Remote Sensing}
  \bibinfo{volume}{138} (\bibinfo{year}{2018}) \bibinfo{pages}{218 -- 231}.
\bibitem[{Hughes et~al.(2018)Hughes, Auer, and Schmitt}]{Hughes2018}
\bibinfo{author}{L.~H. Hughes}, \bibinfo{author}{S.~Auer},
  \bibinfo{author}{M.~Schmitt},
\newblock \bibinfo{title}{Investigation of joint visibility between sar and
  optical images of urban environments},
\newblock \bibinfo{journal}{ISPRS Annals of the Photogrammetry, Remote Sensing
  and Spatial Information Sciences} \bibinfo{volume}{4} (\bibinfo{year}{2018})
  \bibinfo{pages}{129--136}.

\end{thebibliography}







\end{document}